\documentclass[11pt]{article}

\usepackage{amssymb,amsmath,mathrsfs,epstopdf,slashed,color,tikz}
\usetikzlibrary{matrix}
\usetikzlibrary{positioning}
\usepackage{enumitem}
\setlist[itemize]{leftmargin=*}
\usepackage{ifpdf}
\ifpdf
\usepackage{graphicx,placeins,hyperref,cite}
\else
\usepackage[dvipdfmx]{graphicx}
\usepackage[dvipdfmx]{hyperref}
\usepackage[numbers]{natbib}
\usepackage{hypernat}
\fi

\allowdisplaybreaks[1]

\makeatletter

\@addtoreset{equation}{section}
\makeatother

\newcommand{\rd}{{\rm d}}

\definecolor{darkgreen}{rgb}{0.11,0.70,0.25}

\begin{document}

\setcounter{page}{1}
\renewcommand{\thefootnote}{\fnsymbol{footnote}}
\rightline{KOBE-COSMO-22-10}
~
\vspace{.80truecm}

\begin{center}
{\fontsize{21}{18} \bf Stability of Hairy Black Holes in Shift-Symmetric Scalar-Tensor Theories via the Effective Field Theory Approach}
\end{center}

\vspace{1cm}

\begin{center}
{\fontsize{13}{18}\selectfont
Justin Khoury,${}^{\rm a}$\footnote{\href{mailto:jkhoury@sas.upenn.edu}{\texttt{jkhoury@upenn.edu}}}
Toshifumi Noumi,${}^{\rm b}$\footnote{\href{mailto:tnoumi@phys.sci.kobe-u.ac.jp}{\texttt{tnoumi@phys.sci.kobe-u.ac.jp}}}
Mark Trodden,${}^{\rm a}$\footnote{\href{mailto:trodden@upenn.edu}{\texttt{trodden@upenn.edu}}}
and
Sam S. C. Wong${}^{\rm a}$\footnote{\href{mailto:scswong@sas.upenn.edu}{\texttt{scswong@sas.upenn.edu}}}
}
\end{center}
\vspace{0.8cm}

\centerline{{\it ${}^{\rm a}$Center for Particle Cosmology, Department of Physics and Astronomy,}}
\centerline{{\it University of Pennsylvania 209 S. 33rd St., Philadelphia, PA 19104, USA}}
 \vspace{.3cm}

 \centerline{{\it ${}^{\rm b}$Department of Physics, Kobe University, Kobe 657-8501, Japan}}
 
  \vspace{.25cm}

 \vspace{1cm}

\begin{abstract}
\noindent
Shift-symmetric Horndeski theories admit an interesting class of Schwarzschild-de Sitter black hole solutions exhibiting time-dependent scalar hair. The properties of these solutions may be studied via a bottom-up effective field theory (EFT) based on the background symmetries. This is in part possible by making use of a convenient coordinate choice --- Lema\^itre-type coordinates --- in which the profile of the Horndeski scalar field is linear in the relevant time coordinate. We construct this EFT, and use it to understand the stability of hairy black holes in shift-symmetric Horndeski theories, providing a set of constraints that the otherwise-free functions appearing in the Horndeski Lagrangian must satisfy in order to admit stable black hole solutions. The EFT is analyzed in the decoupling limit to understand potential sources of instability. We also perform a complete analysis of the EFT with odd-parity linear perturbations around general spherically symmetric space-time.

\end{abstract}

\newpage

\setcounter{tocdepth}{2}
\tableofcontents

\renewcommand*{\thefootnote}{\arabic{footnote}}
\setcounter{footnote}{0}

\section{Introduction}
In the more than a hundred years since the discovery of black hole solutions to the theory of General Relativity (GR), enormous theoretical efforts have been devoted to understanding these remarkable space-time objects. Nevertheless, it has only been in the last decade that two stunning observational results --- the measurement of gravitational waves from binary mergers by the LIGO collaboration~\cite{TheLIGOScientific:2016pea}, and the imaging of super massive black hole shadows by the Event Horizon Telescope~\cite{Akiyama:2019cqa} --- have provided direct observational evidence for astrophysical black holes. These observations have lifted the curtain on the new era of black hole science, and these and other upcoming precision measurements will provide powerful tools for scrutinizing theoretical ideas beyond our currently established theories. Among these new tools, of particular importance for this paper are the measurements of black hole quasinormal mode frequencies. 

A binary black hole merger proceeds through three somewhat distinct phases; inspiral, merging and ringdown. During the ringdown phase, deviations from the black hole space-time gradually decay, and therefore can be treated as small perturbations around the black hole solution~\cite{Regge:1957td,Zerilli:1970se}. The study of the linear system of these small perturbations around a black hole is sometimes called black hole spectroscopy, since the decaying oscillations exhibit a set of characteristic wave forms called quasinormal modes, with corresponding frequencies known as {\it quasinormal} frequencies \cite{Berti:2009kk}. The quasinormal frequency spectrum of a neutral, spinning black hole in GR is entirely determined by two parameters, the mass and spin of the black hole. Thus, it is theoretically possible to probe proposed modifications to pure GR black holes by studying even just the lowest two ringdown tones. Thus, quasinormal mode analysis provides a particularly promising way to constrain, rule out or even provide hints for theoretical constructions beyond GR, and we expect that the increasingly accurate measurements provided by forthcoming missions, such as LISA, will yield important new insights through this method.

A particularly well-motivated and well-studied class of modifications to GR is scalar-tensor theories, which arise in many different settings. While a general effective field theory (EFT) construction of such theories would include all operators up to a certain order, if one only allows a restricted set of operators involving a scalar and the metric, such that they give rise to second-order equations of motion, then such theories are free of a particularly pernicious instability known the ``Ostragradsky ghost"~\cite{Woodard:2015zca}. (See~\cite{Solomon:2017nlh} for a systematic discussion of the extent to which it is necessary to include higher-derivative operators in the EFT of general scalar-tensor theories, and the circumstances under which it is correct to restrict to only second-order operators.) 

The most general scalar-tensor theory with second-order field equations is known as Horndeski theory~\cite{Horndeski:1974wa}. It was rediscovered fifteen years ago, along with Galileon theories~\cite{Nicolis:2008in},
which were subsequently covariantized to curved space-time~\cite{Deffayet:2009wt,Deffayet:2009mn,Deffayet:2011gz,Kobayashi:2011nu}. It is also possible to study generalizations of Horndeski to 
GLPV~\cite{Gleyzes:2014dya} or DHOST~\cite{Langlois:2015cwa,Crisostomi:2016czh,BenAchour:2016fzp,Takahashi:2017pje,Langlois:2018jdg} theories, in which the equations of motion are nominally third-order, but which can be shown to be reducible to second-order ones. See~\cite{Kobayashi:2019hrl} for a review.

It is well-known that scalar-tensor theories admit numerous {\it hairy} black hole solutions, in which one or more fields take on a nontrivial profile around the black hole, and for which, as a result, quantum numbers beyond the charge, spin, and mass are required to characterize them. One class of hairy solutions admits a radially-dependent scalar profile~$\phi(r)$~\cite{Sotiriou:2013qea,Sotiriou:2014pfa,Babichev:2016rlq,Benkel:2016rlz,Babichev:2017guv,Lehebel:2017fag,Minamitsuji:2018vuw,BenAchour:2019fdf,Minamitsuji:2019tet}. The general quasinormal mode analysis for this type of black hole has been carried out~\cite{Kobayashi:2012kh,Kobayashi:2014wsa} and, remarkably, it has also been shown that the properties of perturbations around black holes with such a radial hair profile can be captured by an EFT description~\cite{Franciolini:2018uyq}. Perhaps more relevant to a cosmological setting, another class of hairy black hole solutions features a time-dependent profile~$\phi(t,r)$, with a time-like gradient~\cite{Babichev:2013cya,Kobayashi:2014eva,Babichev:2016kdt,Babichev:2017lmw,BenAchour:2018dap,Motohashi:2019sen,Takahashi:2019oxz,Minamitsuji:2019shy,Minamitsuji:2019tet,Khoury:2020aya}. These black holes exist in a wide class of scalar-tensor theories satisfying a shift symmetry $\phi\to\phi +c$. They are akin to the time-dependent scalar profile that drives cosmological expansion in ghost condensation~\cite{ArkaniHamed:2003uz,Mukohyama:2005rw}. These time-dependent hairy black hole solutions require careful study, since it has been shown in certain circumstances that they may suffer from a strong coupling problem \cite{Ogawa:2015pea,Babichev:2017lmw,Babichev:2018uiw,deRham:2019gha} and/or a gradient instability~\cite{Khoury:2020aya,Takahashi:2021bml,Minamitsuji:2022mlv,Minamitsuji:2022vbi}. 

In this work we analyze hairy black holes with time-like hair using an EFT approach. The construction of the EFT follows the logic of the EFT of inflation~\cite{Cheung:2007st} and the EFT of the purely radial hairy solution~\cite{Finelli:2018upr}. The first example of such a construction was presented in~\cite{Mukohyama:2022enj}. Here, we further constrain the EFT coefficient functions using the isometries of the background space-time and the shift symmetry of the scalar field, with the primary goal of diagnosing the existence of gradient instabilities. While a complete analysis of tensor fluctuations in the EFT remains highly technical, it can be shown that gradient instabilities that persist in the decoupling limit cannot be cured through mixing with tensor fluctuations away from that limit. Thus, it is possible to obtain useful results by studying the theory in the decoupling limit. We analyze effective operators in the EFT individually to provide constraints and insights about potential instabilities from the choices of coefficient functions.  We also carry out a comprehensive analysis of odd sector perturbations. We further obtain useful constraints on the EFT by considering the stealth black hole limit, in which the background geometry reduces to Schwarzschild-de Sitter, while the scalar profile remains non-trivial.

\section{Background geometry and Ingredients for the EFT}  \label{sec:setup}

We focus on black hole (BH) solutions in general scalar-tensor theories such that the geometry is a static, spherically symmetric solution,
\begin{equation} 
\label{eqn:gtr}
 \rd s^2 = - f(r) \rd t^2 + \frac{\rd r^2}{g(r)} + r^2 \rd \Omega^2\,,
 \end{equation}
with~$\rd\Omega^2=\rd\theta^2+\sin^2\theta \rd\varphi^2$. We adopt Lema\^itre-type coordinates for this type of geometry,
\begin{align} \label{eqn:glemaitre}
  \rd s^2 = -\rd \tau^2 +\big(1-f(r)\big) \rd \rho^2 + r^2 \rd \Omega^2\,,
\end{align}
where \{$\tau,\rho$\} are related to \{$t,r$\} through 
\begin{align}
\rd t  &= \frac{1}{f} \rd \tau  + \left(1 - \frac{1}{f} \right) \rd \rho\,; \nonumber \\
\rd r &=\sqrt{\frac{(1-f)g}{f} }  \big( - \rd \tau + \rd \rho \big)\,.
\end{align}
Throughout this paper we use the notation $(\;)\dot{}  \equiv \frac{\partial}{\partial \tau}(\;)$ and  $(\;)'  \equiv \frac{\partial}{\partial \rho}(\;)$. It is clear from the metric that this coordinate system is synchronous, and constant $(\rho ,\, \Omega)$ are free-falling time-like geodesics. When the theory enjoys a shift symmetry for the scalar, $\phi$, it supports a class of interesting hairy black holes with time-dependent hair of the form\cite{ArkaniHamed:2003uy},\footnote{In general, the gradient of a time-dependent scalar hair does not necessarily align with~$\partial_\tau$. In such a case, it is more convenient to use a time coordinate that aligns with $\nabla_\mu \phi$.}
\begin{align} \label{eqn:phi}
  \bar{\phi} (x) = m^2 \tau\,. 
\end{align}
Furthermore, for the subclass of shift-symmetric theories we are interested in, the action of fluctuations exhibits the set of symmetries inherited from the background isometries, as we will see in Sec.~\ref{Sec:EFT_perturbations}.

The first step in constructing the EFT of fluctuations about this black hole is to note that the gradient of the scalar field,
\begin{equation}
 n^{\mu} = -\frac{\nabla^{\mu}\phi}{\sqrt{-\nabla_{\nu}\phi\nabla^{\nu}\phi }}\,; \quad n^{\mu} n_{\mu}=-1\,,
\end{equation}
defines a foliation of space-time. This gradient is the vector normal to surfaces of constant-$\phi$. From here until Sec.~\ref{Sec:EFT_perturbations}, we work in unitarity gauge defined by $\phi=\bar{\phi}$, in which these surfaces reduce to constant-$\tau$ surfaces, denoted by $\Sigma_{\tau}$. From this one can define an induced metric on such hypersurfaces, 
\begin{equation}
h_{\mu\nu}=g_{\mu\nu}+n_{\mu}n_{\nu}\, ,
\end{equation}
which also serves as a projector onto these surfaces. For example, the covariant derivative on the surface $D_{\mu}$ is defined by 
\begin{align}
 D_{\mu} V_{\nu} = h^{\alpha}_{\;\mu} h^{\beta}_{\;\nu} \nabla_{\alpha} V_{\beta}\, ,
\end{align}
and the extrinsic curvature of surfaces is given by 
\begin{align}
 K_{\mu\nu} = h_{\;\mu}^{\alpha}h_{\;\nu}^{\beta}\nabla_{\alpha}n_{\beta}= h_{\;\mu}^{\alpha}\nabla_{\alpha} n_{\nu} = \nabla_{\mu}n_{\nu} + n_{\mu}n^{\alpha}\nabla_{\alpha} n_{\nu} \, .
\end{align}

Using the ADM form of the metric,
\begin{align}
\rd s^2 = -N^2 \rd \tau^2 + h_{ij}\left( \rd x^i+N^i \rd \tau\right)\left( \rd x^j +N^j\rd \tau\right) \, ,
\end{align}
this can be written more explicitly in component form, 
\begin{align}
K^{\tau\nu} =0\,; \quad K_{ij} = \frac{1}{2N}\left( \dot{h}_{ij} - D_i N_j - D_j N_i \right)\,,
\end{align}
where we have written $n^{\mu}$ in terms of the lapse function $N$ and the shift vector $N_i$,
\begin{equation}
  n^{\mu} = \left( \frac{1}{N} , -\frac{N^i}{N} \right), \quad n_{\mu} = \left( -N ,\vec{0}\right)\,.
\end{equation}
Another important geometric quantity is the intrinsic curvature of the hypersurface,~$\hat{R}_{\mu\nu\rho \sigma}$, which is defined by 
\begin{align}
 \hat{R}_{\mu\nu\rho}^{\quad\;\; \sigma} V_{\sigma} = [D_{\mu},D_{\nu}] V_{\rho}\, .
\end{align}
Together with $K_{\mu\nu}$, this is related to the Riemann tensor of the full space-time via the Gauss-Codazzi relations
\begin{align}
 \hat{R}_{\mu\nu\rho}^{\quad\;\; \sigma} + K_{\mu \rho} K_{\nu}^{\;\;\sigma} - K_{\nu\rho}K_{\mu}^{\;\; \sigma} & = h_{\mu}^{\;\;\alpha}h_{\nu}^{\;\;\beta}h_{\rho}^{\;\;\gamma}h^{\sigma}_{\;\;\delta}R_{\alpha \beta \gamma}^{\quad \;\; \delta} \,; \\
D_{\mu}K^{\mu}_{\;\;\nu} - D_{\nu} K &= R_{\alpha \beta} n^{\alpha} h^{\beta}_{\;\; \nu}  \,.
\end{align}

Evaluated on the background~\eqref{eqn:glemaitre}, the above geometric quantities are 
\begin{align} \label{eqn:bgK}
\bar{K}^{\rho}_{\;\; \rho} &= \frac{f_r}{2(1-f)}\sqrt{\frac{g}{f} }\,; \nonumber \\
\bar{K}^{\theta}_{\;\; \theta}& = \bar{K}^{\varphi}_{\;\; \varphi}  = -\frac{1}{r} \sqrt{\frac{(1-f)g}{f}}\,; \\
\bar{\hat{R}}^{\rho}_{\;\; \rho} &= \frac{g f_r - f g_r}{r f^2 } \,;\nonumber \\
\bar{\hat{R}}^{\theta}_{\;\; \theta}& = \bar{\hat{R}}^{\varphi}_{\;\; \varphi}  = \frac{2 f^2 -2fg + r gf_r - r f g_r }{2 r^2 f^2 }\,,
\end{align}
where $(\;)_r = \frac{\partial}{\partial r}(\;)$. One can then see that the traceless part of $\bar{K}_{ij}$ and $\bar{\hat{R}}_{ij}$,
\begin{align}
 \bar{K}^{\rm T}{}^{i}_{\;\;j}  &= \bar{K}{}^{i}_{\;\;j}  - \frac{1}{3} \delta^{i}_{\;\;j} \bar{K},  \nonumber \\
  \bar{\hat{R}}^{\rm T}{}^{i}_{\;\;j} &= \bar{\hat{R}}{}^{i}_{\;\;j}  - \frac{1}{3} \delta^{i}_{\;\;j} \bar{\hat{R}}, 
\end{align}
 are proportional to each other, $\bar{\hat{R}}^{\rm T}{}^{i}_{\;\;j} \propto \bar{K}^{\rm T}{}^{i}_{\;\;j} $. Note that when $f(r) = g(r)$, all components of $\bar{\hat{R}}_{\mu\nu\rho}^{\quad\;\; \sigma}$ on $\Sigma_{\tau}$ vanish. With all of these three-dimensional covariant quantities at hand, we are now ready to build the EFT. 

\section{The effective action for perturbations}
\label{Sec:EFT_perturbations}

The most general action for perturbations, including all terms up to second order in fields and second derivatives that describe scalar tensor theories of our interest, is given by 
\begin{align} \label{eqn:EFT}
 S_{(2)} = \int \rd ^4x \sqrt{-g}\bigg[& \frac{M_{1}(x)}{2}R - \Lambda(x) + \alpha(x) g^{\tau\tau}   + \beta(x) \bar{K}^{\nu}_{\;\;\mu}K^{\mu}_{\;\;\nu} \nonumber \\
 &+ M_2(x) (\delta g^{\tau\tau})^2   + M_3(x) \delta g^{\tau\tau} \delta K + M_4(x)\bar{K}_{\;\;\mu}^{\nu} \delta g^{\tau\tau} \delta K^{\mu}_{\;\;\nu} \nonumber \\
 &+ M_5(x) (\partial_{\tau}\delta g^{\tau\tau})^2 + M_6(x)(\partial_{\tau}\delta g^{\tau\tau}) \delta K  + M_7(x) \bar{K}_{\;\;\mu}^{\nu}(\partial_{\tau}\delta g^{\tau\tau}) \delta K^{\mu}_{\;\;\nu}+ M_8(x) (D_i \delta g^{\tau\tau })^2   \nonumber \\
 & + M_9(x) \delta K^2  + M_{10}(x) \delta K_{\;\;\mu}^{\nu}\delta K^{\mu}_{\;\;\nu} + M_{11}(x) \bar{K}_{\;\;\mu}^{\nu}\delta K \delta K^{\mu}_{\;\;\nu} + M_{12}(x)\bar{K}_{\;\;\mu}^{\nu} \delta K_{\;\;\mu}^{\rho}\delta K_{\;\;\rho}^{\nu} \nonumber \\
& + M_{13}(x)\delta g^{\tau\tau} \delta \hat{R} + M_{14}(x)  \bar{K}_{\;\;\mu}^{\nu}\delta g^{\tau\tau} \delta \hat{R}^{\mu}_{\;\;\nu} \bigg] \, ,
\end{align}
where $x\equiv \{\tau,\,\rho\}$. As we will soon show, the coefficient functions $\{\alpha(x), \,\beta(x) ,\,\dots,\, M_i(x),\,\dots\} $ will reduce to functions of $r(\tau,\rho)$ when describing perturbations around \eqref{eqn:gtr}. However, unless otherwise specified, in the following sections we will use the slightly more general action above. 

We are of course free to perform the analysis in any frame, such as Jordan frame or Einstein frame, by performing a suitable conformal transformation. Such transformation would only affect the coupling of~$\phi$ to matter fields, which we ignore in our analysis. Without loss of generality, therefore, we choose to work in Einstein frame, setting~$M_1(x)$ to be constant and equal to the reduced Planck mass:~$M_1(r) = M_{\rm Pl}^2 \equiv \frac{1}{8\pi G}$.  

The above action can in principle be derived from the expansion in perturbations of a general function in terms of the building blocks of the EFT,
\begin{align} \label{eqn:masteraction}
  S = \int\rd^4x \sqrt{-g}\, {\cal L}(g^{\mu\nu}, R_{\mu\nu\rho \sigma}, g^{\tau\tau}, K_{\mu\nu}, \nabla_{\mu}; \tau ) \,.
\end{align}
In the following we will explain the construction of tadpoles and quadratic terms, and their relation to the most general action~\eqref{eqn:masteraction}.

\subsection{Tadpole terms}
The tadpole terms of the action are constructed from the building blocks $\delta g^{\tau\tau}$, $K_{\mu\nu}$, $R_{\mu\nu\rho \sigma}$ and scalar functions of the background coordinates. Their most general form is 
\begin{align}
 S_{\rm tad.} = \int \rd^4x \sqrt{-g}\,\Big[ \Lambda(x) +\alpha(x) g^{\tau\tau} + \beta_{\;\;\mu}^{\nu}(x)K^{\mu}_{\;\;\nu} + \zeta_{\mu\nu}^{\quad \rho\sigma}(x) R^{\mu\nu}_{\quad \rho\sigma}  \Big]\,. 
\end{align}
By construction, only the components $K^{j}_{\;\;i}$ are non-zero, and hence the third term reduces to~$\beta^{i}_{\;\;j}K_{\;\;i}^{j}$, where $\beta^{i}_{\;\;j}$ is constructed from the set of tensor functions that obey the background isometries, namely $\bar{g}^{i}_{\;\;j}$ and $\bar{K}^{i}_{\;\;j}$.\footnote{Since the 2-sphere is maximally-symmetric, all other tensors compatible with the background isometries should be linear combinations of~$\bar{g}_{ij}$ and~$\bar{K}_{ij}$. We have seen in \eqref{eqn:bgK} that $\bar{\hat{R}}_{ij}$ is an example.} Note also that, obviously,~$K = \nabla_\mu n^{\mu}$ is a total derivative term. Thus the two index tensor $\beta^{\mu}_{\;;\nu}$ can be chosen as $\beta(x) \bar{K}^{\mu}_{\;\;\nu}$. We refer the reader to~\cite{Cheung:2007st,Finelli:2018upr} for the detailed construction of the term~$\zeta_{\mu\nu}^{\quad\rho\sigma}(x) R^{\mu\nu}_{\quad \rho\sigma} $. The tadpole action gives rise to background equations of motion for $g^{\mu\nu}$ as shown in Appendix~\ref{app:bgEOM}. When applied to the ansatz \eqref{eqn:glemaitre}, it generates a set of consistency relations between the coefficient functions $M_1$, $\Lambda$, $\alpha$ and  $\beta$ and the background metric.

\subsection{Quadratic terms}
It is useful to start the analysis by thinking of possible terms in the action \eqref{eqn:masteraction}. There are infinitely many contractions of~$K_{\mu\nu}$ that obey diffeomorphism invariance on $\Sigma$, for instance 
\begin{align}
 &a_1(g^{\tau\tau}, \partial g^{\tau\tau})(K)^{n_1} \,,  \quad  a_2(g^{\tau\tau}, \partial g^{\tau\tau})(K)^{n_2} K_{\;\;\mu}^{\nu}K^{\mu}_{\;\;\nu}\,, \quad a_3(g^{\tau\tau}, \partial g^{\tau\tau})K_{\;\;\mu}^{\nu} K^{\mu}_{\;\;\rho}K^{\rho}_{\;\;\nu}\,, \nonumber \\
& \quad a_4(g^{\tau\tau}, \partial g^{\tau\tau}) \hat{R}\,, \quad  a_5(g^{\tau\tau}, \partial g^{\tau\tau}) K \hat{R}\,, \quad  a_6(g^{\tau\tau}, \partial g^{\tau\tau}) K_{\;\;\mu}^{\nu} \hat{R}^{\mu}_{\;\;\nu} \,,~\ldots 
\end{align}
where the $a_i$ are arbitrary functions of~$g^{\tau\tau}$, and where $n_1= 0, 1,\, 2,\, 3$ and $ n_2 =0, 1$.

Since we have in mind that the EFT describes perturbations of Horndeski or DHOST theories, the maximal number of derivatives in each operator is three~\cite{Gleyzes:2013ooa}. For example, the operators $K^3$, $K_{\;\;\mu}^{\nu}K^{\mu}_{\;\;\nu}$ and $K_{\;\;\mu}^{\nu}K^{\nu}_{\;\;\rho} K^{\rho}_{\;\;\mu}$ originate from the Horndeski~$G_5$ or cubic DHOST term. Therefore, the above set of operators are all the possible operators involving $K_{\;\;\mu}^{\nu}$ and $\hat{R}^{\mu\nu}_{\quad \rho\sigma}$. Expanding these up to and quadratic order in perturbations~$\delta g^{\tau\tau}= g^{\tau\tau} - \bar{g}^{\tau\tau} $, $\delta K = K - \bar{K}$, $\delta K_{\;\;\mu}^{\nu} = K_{\;\;\mu}^{\nu} - \bar{K}_{\;\;\mu}^{\nu}$, $\delta \hat{R} = \hat{R}  - \bar{\hat{R} }$ and $\delta  \hat{R}_{\;\;\mu}^{\nu} =  \hat{R}_{\;\;\mu}^{\nu} - \bar{ \hat{R}}_{\;\;\mu}^{\nu}$, and up to second-order in derivatives, we arrive at the quadratic terms in \eqref{eqn:EFT}. 

Note that when one specializes to the EFT of perturbations around the background~\eqref{eqn:glemaitre} with $n^{\nu}$ defined by~\eqref{eqn:glemaitre},  the quadratic terms in \eqref{eqn:EFT} can actually describe perturbations of additional operators such as $K_{\;\;\;\nu_1}^{ \mu_1}K^{\nu_1}_{\;\;\; \nu_2} \dots K^{\nu_n}_{\;\;\; \mu_1}$, since $\bar{K}_{\;\;\;\nu_1}^{ \mu_1}\bar{K}^{\nu_1}_{\;\;\; \nu_2} \dots \bar{K}^{\nu_{n-1}}_{\quad \nu_n}$ can be written as a linear combination of~$\bar{h}_{\;\;\; \nu_n}^{\mu_1 }$ and~$\bar{K}_{\;\;\;\nu_n}^{\mu_1}$. In principle, if one seeks to describe an arbitrary modified gravity theory, more independent operators (up to second-order in derivatives) can be added to the action, such as $(\bar{K}^{\mu}_{\;\;\nu}\delta K_{\;\;\mu}^{\nu})^2$ and $\bar{K}^{\mu}_{\;\;\nu} \bar{K}^{\alpha}_{\;\; \beta} \delta K^{\nu}_{\;\;\alpha}\delta K_{\;\; \mu}^{\beta}$.


\subsection{Consistency with diffeomorphisms on the surface $\Sigma_{\tau}$}

Since the operators~$\delta K^{\mu}_{\;\;\nu}$ and~$\delta \hat{R}^{\mu\nu}_{\quad \rho\sigma}$ do not transform covariantly under diffeomorphisms on~$\Sigma_{\tau}$, the above action for perturbations, with general coefficients, explicitly breaks diffeomorphism invariance on $\Sigma_{\tau}$. However, as mentioned in~\cite{Franciolini:2018uyq, Mukohyama:2022enj}, the EFT~\eqref{eqn:EFT} for perturbations derives from the parent action~\eqref{eqn:masteraction}, which is constructed to be invariant under {\it all} symmetries. As such, the action for perturbations should respect diffeomorphisms on $\Sigma_{\tau}$. More explicitly, it has been shown in~\cite{Mukohyama:2022enj} that these two observations are reconciled by diffeomorphism invariance of the action implying a set of non-trivial constraints on the EFT coefficient functions in~\eqref{eqn:EFT}.  

\subsection{Background isometry and shift symmetry} 
\label{sec:isometry}

The background~\eqref{eqn:glemaitre} has four Killing vector fields, 
\begin{align} \label{eqn:killvec}
 v_1 &= \partial_t = \partial_{\tau}+ \partial_{\rho}\,; \nonumber \\
 J_1 &=  - \sin \varphi \partial_{\theta}  - \cot \theta \cos \varphi \partial_{\varphi}   \,; \nonumber \\
 J_2 &=  \cos \varphi \partial_{\theta}  - \cot \theta \sin \varphi \partial_{\varphi}   \,; \nonumber \\
 J_3 &= \partial_{\varphi} \,, 
\end{align}
associated with the time translation and rotation invariances of the black hole background. 
The coefficient functions in the perturbation Lagrangian are obtained by expanding~\eqref{eqn:masteraction}. For example:
\begin{align}
\left. \frac{\partial^2 {\cal L} }{(\partial g^{\tau\tau})^2 }\right|_{\bar{g}_{\mu\nu} ,\bar{n}^{\mu}} (\delta g^{\tau \tau })^2 + \left. \frac{\partial^2 {\cal L} }{\partial g^{\tau\tau} \partial K }\right|_{\bar{g}_{\mu\nu} ,\bar{n}^{\mu}} \delta g^{\tau \tau }\delta K +  \left. \frac{\partial^2 {\cal L} }{\partial g^{\tau\tau} \partial K^{\mu\nu} }\right|_{\bar{g}_{\mu\nu} ,\bar{n}^{\mu}} \delta g^{\tau \tau }\delta K^{\mu\nu} +\ldots 
\end{align}
Since all the coefficients~$ \left.\frac{\partial^2 {\cal L} }{(\partial \dots  )^2 }\right|_{\bar{g}_{\mu\nu}, \bar{n}^{\mu}}$ are evaluated on the background, and since the action~${\cal L}$ is itself constructed to respect the symmetries, the coefficients must respect all the isometries generated by~\eqref{eqn:killvec}. Therefore they should obey 
\begin{align}
   {\cal L}_{v_1} \left.\frac{\partial^2 {\cal L} }{(\partial \dots  )^2 }\right|_{\bar{g}_{\mu\nu}, \bar{n}^{\mu}} =  {\cal L}_{J_1} \left.\frac{\partial^2 {\cal L} }{(\partial \dots  )^2 }\right|_{\bar{g}_{\mu\nu}, \bar{n}^{\mu}}  =  {\cal L}_{J_2} \left.\frac{\partial^2 {\cal L} }{(\partial \dots  )^2 }\right|_{\bar{g}_{\mu\nu}, \bar{n}^{\mu}} =  {\cal L}_{J_3} \left.\frac{\partial^2 {\cal L} }{(\partial \dots  )^2 }\right|_{\bar{g}_{\mu\nu}, \bar{n}^{\mu}}  =0 \,,
\end{align}
where ${\cal L}_{X}$ denotes the Lie derivative. As a result, as we mentioned earlier, we must demand that all the scalar coefficients $\{\alpha(x), \,\beta(x) ,\,\dots,\, M_i(x),\,\dots\} $ are {\it functions of $r$ only}, and that any background tensor of the form $\bar{T}_{AB}$ {\it must be proportional to $\gamma_{AB}$}, where~$\gamma_{AB}$ is the metric on the 2-sphere, and~$A,B$ are coordinate indices on the sphere. 

However, we must take care here, since the time-dependent scalar profile $\bar{\phi} = m^2\tau$ spontaneously breaks the isometry generated by $v_1$. Nevertheless, in addition to the isometries, there is an additional shift symmetry, $\phi \to \phi  +c$, in the class of scalar-tensor theories of interest. Because of this, the isometry generated by $v_1$ remains a good symmetry in the action for perturbations. In other words, in the particular case we are interested in, due to the existence of the time isometry, {\it the shift symmetry is equivalent to an isometry.} To see this, we write the background scalar profile as 
\begin{align}
 \bar{\phi} = m^2\tau = \Lambda^2(t + \psi(r) )\,.
\end{align}
A shift in $\bar{\phi} $, due to transforming $\phi \to \phi + c$,  can therefore be absorbed into~$t$ by making use of an isometry, $t \to t-\frac{c}{m^2}$. In other words, the spontaneously broken time translation due to $\bar{\phi}$ is absorbed by a shift in $\phi$, while the ``{\it diagonal}" part of the time translation and $\phi$-shift symmetry remains unbroken. \footnote{For instance,  it was demonstrated  in \cite{Akhoury:2008nn} that stationary configurations requires shift symmetry in k-essence theories.}  Only when the scalar profile is more complicated, in such a way that the diffeomorphism performed to absorb the shift in $\phi$ cannot be an isometry, will this lead to more non-trivial constraints on the EFT coefficients~\cite{Finelli:2018upr}.\footnote{An explicit example is the ultra slow-roll regime, where~$\bar{\phi} \approx \frac{c}{\tau ^6}$ in conformal time $\tau$.}

\section{Second order action in the decoupling limit}

Previous studies of a class of black holes with time-dependent hair in scalar-tensor theories have shown that gradient instabilities are inevitable, since the radial and angular sound speed squared of perturbations have opposite signs~\cite{Khoury:2020aya,Takahashi:2021bml}. We have also learned that it is the scalar modes that are responsible for this instability~\cite{Khoury:2020aya}. While establishing the stability of a theory against perturbations requires a complete analysis, proving the existence of an instability requires much less work. 

For this purpose, it is practically useful to analyze stability in the decoupling limit. In general, our EFT has two characteristic scales. One is the decoupling scale $\Lambda_{\rm dec}$, at which gravity decouples from the Nambu-Goldstone mode $\pi$. For example, consider the operators $\frac{M_{\rm Pl}^2}{2}R + \alpha(\tau,\rho) g^{\tau\tau}$ under Stuckelberg's transformation,
\begin{align}
&\quad \frac{M_{\rm Pl}^2}{2}R+ \alpha(\tau,\rho) g^{\tau\tau} \nonumber \\
&=\frac{M_{\rm Pl}^2}{2}R + \alpha(\tau+\pi,\rho) \Big(g^{\tau\tau}(1+\dot{\pi})^2 + 2 g^{\tau a}\partial_a(1+\dot{\pi}) + g^{ab} \partial_a \pi \partial_b \pi \Big) 
\end{align} 
After introducing the canonically normalized fields, $\pi =\frac{\pi_c}{\sqrt{2\alpha}} $ and $g^{\mu\nu} = \bar{g}^{\mu\nu} + \frac{1}{M_{\rm p}}\delta g_c^{\mu\nu}$, the mixing term $ \frac{\sqrt{2\alpha}}{M_{\rm Pl}} \delta g^{\tau\tau}_c \dot{\pi}_c $ indicates that the scale of mixing is $E_{\rm mix} = \frac{\sqrt{2\alpha}}{M_{\rm Pl}}$. In the limit $M_{\rm Pl} \to \infty$ with $\sqrt{2\alpha}$ fixed, the $\pi$ action decouples from the metric fluctuations, which analogous to the EFT of inflation~\cite{Cheung:2007st}. We focus on the region of parameter space such that $E_{\rm mix} \ll \Lambda_{\rm dec}$.

The other is the cutoff scale $\Lambda_{\rm EFT}$, beyond which the EFT is not trustworthy, either due to strong coupling or a breakdown of the derivative expansion. One may study the dispersion relation of $\pi$ up to the second order derivatives. If a gradient instability appears at some scale $\Lambda_{\rm grad}$ in the regime $\Lambda_{\rm dec}<\Lambda_{\rm grad}<\Lambda_{\rm EFT}$, it indicates an instability. Recall that $\Lambda_{\rm EFT}$ depends on details of higher order terms, which is beyond the scope our present work. Thus, we study (in)stability in the decoupling limit based on the dispersion relation up to second-order derivatives.


One way to study the decoupling limit in the EFT language is to introduce a St\"uckelberg field $\pi(x^{\mu})$ to explicitly restore time-translational invariance in~$\tau$~\cite{Cheung:2007st}. Under $\tau \to \tilde{\tau} = \tau + \pi(x^{\mu})$, the important elements are
\begin{align}
	 \left[ \frac{\partial \tilde{x}^{\mu}(x) }{\partial x^{\nu}}\right] &= \begin{pmatrix}
	1+ \dot{\pi}(t,\vec x) &  \partial_k \pi(t,\vec{x}) \\
	\mathbf{0} & \delta^{i}_{\;j}
	\end{pmatrix}\,;  \nonumber \\
	\left[ \frac{\partial x^{\mu}(\tilde{x}) }{\partial \tilde{x}^{\nu}} \right] &= \begin{pmatrix}
	\frac{1}{1+ \dot{\pi}(t,\vec x)} &  -\frac{\partial_k \pi(t,\vec{x})}{1+ \dot{\pi}(t,\vec x)}\\
	\mathbf{0} & \delta^{i}_{\;j}
	\end{pmatrix}  \,,
\end{align}
where~$\mu$ and~$\nu$ are row and column indices, respectively. For instance, the transformation~$\tilde{g}_{\mu\nu}(\tilde{x}) = \frac{\partial x^{\alpha}(\tilde{x})}{ \partial \tilde{x}^{\mu}} \frac{\partial x^{\beta}(\tilde{x})}{ \partial \tilde{x}^{\nu}}g_{\alpha\beta}(x)$ leads to 
\begin{align}
 g_{\tau\tau} &\to g_{\tau\tau}\frac{1}{(1 +\dot{\pi})^2} \,; \nonumber\\
 g_{  \tau i} & \to g_{\tau i}\frac{1}{1 +\dot{\pi}}  - g_{\tau \tau} \frac{\partial_i \pi}{(1 +\dot{\pi})^2}   \,; \nonumber \\
 g_{ij} &\to g_{ij}- g_{\tau i} \frac{\partial_j \pi}{1+ \dot{\pi}}-   g_{\tau j} \frac{\partial_i \pi}{1+ \dot{\pi}} + g_{\tau\tau} \frac{\partial_i \pi\partial_j \pi}{(1 +\dot{\pi})^2} \,.
\end{align}
The transformation rule for the corresponding components in the ADM decomposition is then
\begin{align}
 h_{ij} \to  h_{ij}  -(1-\dot{\pi})\Big(N_i\partial_j \pi + N_j\partial_i \pi \Big) + \left(-N^2 +N^kN_k\right) \partial_i\pi \partial_j\pi  + {\cal O}(\pi^3) \,,
\end{align}
while the spatial Christoffel connection~$\hat{\Gamma}^{l}_{ij} = \frac{1}{2} h^{lk} \left( \partial_{i} h_{jk} + \partial_{j} h_{ik} - \partial_{k} h_{ij}  \right)$ used to define $D_{i}$ transforms as 
\begin{equation}
 \hat{\Gamma}^{l}_{ij}  \to \hat{\Gamma}^{l}_{ij}  - \frac{1}{2}h^{lk}\left(\dot{h}_{ik} \partial_j \pi + \dot{h}_{jk}\partial_i \pi - \dot{h}_{ij} \partial_k \pi \right) + {\cal O}(\pi^2)\,.
\end{equation}
Here we have used that~$\bar{N}_i=0$ on the background geometry. Therefore the relevant transformations for~$\delta K_{ij}$,~$\delta K$,~$\delta \hat{R}_{ij}$ and~$\delta \hat{R}$, up to first order in~$\pi$, become
\begin{align} \label{eqn:curvpi}
  \delta K_{ij} & \to \delta K_{ij} - \dot{\bar{K}}_{ij} \pi  - N D_i D_j \pi\,; \nonumber \\
  \delta K & \to \delta K - \dot{\bar{K}}\pi  - N h^{ij} D_i D_j \pi\,; \nonumber \\
  \delta \hat{R}_{ij} & \to \delta \hat{R}_{ij} -\dot{\bar{\hat{R}}}_{ij} \pi   -\partial_k\pi \dot{\hat{\Gamma}}^k_{ij}  -\frac{1}{2}h^{kl} \dot{\hat{\Gamma}}^m_{kl}  \left(\partial_i \pi h_{mj} +\partial_j\pi h_{mi}   \right) \nonumber \\  
  &\quad\, +\frac{1}{2}\left(\partial_i\pi\dot{\hat{\Gamma}}^m_{jm} +\partial_j\pi\dot{\hat{\Gamma}}^m_{im} + \partial^k\pi \left(\dot{\hat{\Gamma}}^m_{ik}h_{mj} + \dot{\hat{\Gamma}}^m_{jk}h_{mi}\right) \right ) \nonumber \\
&\quad\, - \frac{1}{2} \left( \dot{h}_{li} D^l D_j \pi+\dot{h}_{lj} D^l D_i \pi - h^{kl}\dot{h}_{kl} D_i D_j \pi -\dot{h}_{ij} \hat{\square} \pi \right) ; \nonumber \\
  \delta\hat{R} & \to \delta  \hat{R}- \dot{\bar{\hat{R}}}\pi -2 \partial_k\pi h^{ij}\dot{\hat{\Gamma}}^k_{ij}  + 2 \partial^i \pi \dot{\hat{\Gamma}}^m_{im} - h^{ij} \dot{h}_{li}D^lD_j \pi + h^{kl}\dot{h}_{kl} \hat{\square} \pi \,.
\end{align}

\subsection{Gradient (in-)stability for $\pi$} \label{sec:pistability}

With the scalar mode restored through the St\"uckelberg trick, we are ready to study the stability of the theory in the decoupling limit. Given the symmetries of the background, there are in general two different sound speeds in the problem: the radial sound speed squared, $c_{\rho}^2$, and the angular sound speed squared, $c_{\theta,\varphi}^2$. In the following we individually analyze quadratic operators in the effective action~\eqref{eqn:masteraction}. In each case, we neglect terms that are cubic and higher-order in~$\pi$, as well as terms that involve mixing with~$\delta h_{ij}$. We also neglect terms that are irrelevant to the dispersion relation up to second-order in derivatives, such as $M_{5,6,7,8}$ as they contribute to higher order derivative terms for $\pi$. Note that we treat the coefficients as general functions of~$(\tau, \rho)$, except when it becomes convenient to use the specific dependence of $r(\tau,\rho)$.

\begin{itemize}
\item \underline{$M_2(\tau,\rho)\left(\delta g^{\tau\tau}\right)^2$}: Using the background quantities $\bar{g}^{\tau\tau}=-1$ and $\bar{g}^{a\tau} =0$, this term only contributes to the time-derivative part of $\pi$,
\begin{align}
 \int \rd^4x \sqrt{-g} \,M_2\left(\delta g^{\tau\tau}\right)^2 \to \int \rd^4x \sqrt{-g} \,4 M_2 \dot{\pi}^2 \,. 
\end{align}
In the Schwarzschild-de Sitter limit, $\alpha = \beta =0$, therefore the above term is crucial for generating a $\dot{\pi}^2$ term and $M_2$ must be positive. 
\item \underline{$M_3(\tau,\rho)\, \delta g^{\tau\tau} \delta K$}: The contribution to the~$\pi$ action coming from this term is 
\begin{align}
\int \rd^4x \sqrt{-g}\, M_3\delta g^{\tau\tau} \delta K \to \int \rd^4x \sqrt{-g}\Big[ -(M_3 h^{ij})\dot{} \, D_i \pi D_j \pi + 2 (D^iM_3) \dot{\pi} D_i \pi - m_3(\tau,\rho) \pi^2 \Big]  \,.
\end{align}
Here,~$m_3(\tau,\rho)$ is the mass term, whose explicit form is not necessary for our purposes. (Henceforth,~$m_i(\tau,\rho)$ will denote the mass contribution from the operator with corresponding coefficient~$M_i(\tau,\rho)$.) In general, due to the existence of the mixing term~$2(D^iM_3) \dot{\pi} D_i \pi$, it is necessary to find a coordinate system that diagonalizes the kinetic matrix. In the specific case in which $M_3$ is a function of $r(\tau,\rho)$ we have $\partial_\rho M_3(r) = - \dot{M_3}(r) $, so that the kinetic terms become
\begin{align}
- 2 \dot{M_3}g^{\rho\rho} \dot{\pi} \pi' -(M_3 g^{\rho\rho})\dot{} \,\pi'^2 - \partial_{\tau}\left( \frac{M_3}{r^2}\right) \gamma^{AB} \partial_A \pi \partial_B \pi \,.
\end{align}
Stability along the angular directions therefore requires
\begin{align}
\partial_{\tau}\left( \frac{M_3}{r^2}\right) >0\,.
\end{align}
Given that the~$\dot{\pi}^2$ term (from other operators) is of the form~$A |g^{\tau\tau}| \dot{\pi}^2$, with~$A>0$, a straight forward Hamiltonian analysis \footnote{For a Lagrangian given by
\begin{align}
 L = a  \dot{\pi}^2 + 2 b \dot{\pi} \pi' - c\pi'^2\, ,
\end{align}
the corresponding Hamiltonian is 
\begin{align}
 H = \frac{(P_{\pi} - 2 b \pi' )^2}{4 a } + c \pi'^2 ,\quad  P_{\pi}  = \frac{\partial L}{\partial \dot{\pi}} = 2a \dot{\pi} +2 b \pi' . 
\end{align}
Therefore, the stability conditions are
\begin{align}
 a >0, \quad  c > 0.
\end{align}
} give the following stability constraints:
\begin{align}
(M_3 g^{\rho\rho})\dot{}  \, > 0. 
\end{align}

There also exist suitable coordinates~$(T, \,R)$ that diagonalize the kinetic matrix, with the corresponding kinetic terms
\begin{align}
  \frac{1}{2}\Big( A|g^{\tau\tau}|(2+\delta_3) + \left( M_3g^{\rho\rho}\right)\dot{} \, \delta_3   \Big) (\partial_T \pi)^2 & -  \frac{1}{2}\Big( A|g^{\tau\tau}|\delta_3 +  \left( M_3g^{\rho\rho} \right) \dot{} \, (2+\delta_3) \Big)(\partial_R \pi)^2  \nonumber \\
  &-\partial_{\tau}\left(\frac{M_3}{r^2} \right) \gamma^{AB}\partial_A \pi \partial_B \pi \,,
\end{align} 
where
\begin{equation*}
\delta_3 \equiv \left[1 + \frac{4 \left( \dot{M_3}g^{\rho\rho}\right)^2}{\left(A|g^{\tau\tau}| + \partial_{\tau}\left( M_3g^{\rho\rho} \right) \right)^2}\right]^{\frac{1}{2}} -1 \ge 0 \, .
\end{equation*}

\item \underline{$M_4(\tau,\rho) \bar{K}^{\nu}_{\;\; \mu}\delta g^{\tau\tau} \delta K^{\mu}_{\;\;\nu}$}: This operator generates the following contribution:
\begin{align}
\int \rd^4x \sqrt{-g}\, M_4 \bar{K}^{\nu}_{\;\; \mu}\delta g^{\tau\tau} \delta K^{\mu}_{\;\;\nu} \to \int \rd^4x \sqrt{-g}\Big[- \partial_{\tau}\big( M_4\bar{K}^{ij} \big)  D_i \pi D_j\pi + 2 D_i \big(M_4\bar{K}^{ij}\big) \dot{\pi} D_j \pi - m_4(\tau,\rho) \pi^2 \Big]\,.
\end{align}
Since~$M_4$ is a function of~$(\tau, \rho)$, and using~\eqref{eqn:bgK}, there is no~$\dot{\pi} \partial_A \pi$ mixing. Therefore, without further specifying the geometry, stability along the angular direction requires 
\begin{align}
   \partial_{\tau}\left( M_4\frac{\partial_{\tau}(r^2)}{r^4} \right)>0\,.
\end{align}
As mentioned earlier, other contributions to the~$\dot{\pi}^2$ term are of the form~$A |g^{\tau\tau}| \dot{\pi}^2$, with~$A>0$. Stability obtained from a Hamiltonian analysis requires that 
\begin{align}
 \partial_{\tau}\big( M_4\bar{K}^{\rho\rho} \big)  > 0.
\end{align}

It follows that the two eigenvalues of the kinetic matrix in the~$(\tau,\rho)$ directions are given by
\begin{align}
\lambda_1 = \frac{1}{2} \Big( A |g^{\tau\tau}|  ( 2+ \sigma) + \partial_{\tau} \big(M_4 \bar{K}^{\rho\rho}\big)  \sigma  \Big) \,; \quad \lambda_2 = -\frac{1}{2} \Big( \partial_{\tau} \big(M_4 \bar{K}^{\rho\rho}\big) ( 2+ \sigma) + A |g^{\tau\tau}|   \sigma  \Big) \,,
\end{align}
where 
\begin{equation*}
\sigma \equiv \left[1+ \frac{4 \left( D_i \big(M_4\bar{K}^{i \rho }\big) \right)^2}{\left(A |g^{\tau\tau}| + \partial_{\tau}(M_4 \bar{K}^{\rho\rho} )\right)^2}\right]^{\frac{1}{2}} -1 \ge 0 \, .
\end{equation*}
 
\item \underline{$M_9(\tau, \rho)\delta K^2 + M_{10}(\tau,\rho) \delta K^{\nu}_{\;\; \mu} \delta K^{\mu}_{\;\; \nu}$}: For these operators, it is sufficient to restrict our analysis to the combination 
\begin{align}
   M_9 \left(\delta K^2 - \delta K^{\nu}_{\;\; \mu} \delta K^{\mu}_{\;\; \nu}\right) \,,
\end{align}
since any other choice of relative coefficients differs from this choice only by higher-derivative terms in~$\pi$. We then have 
\begin{align}
&\int \rd^4x \sqrt{-g}\, M_9\left(\delta K^2 - \delta K^{i}_{\;\; j} \delta K^{j}_{\;\; i} \right)  \nonumber \\
&~~~\rightarrow \int \rd^4x \sqrt{-g}\bigg[ \left(M_9 \hat{R}^{ij} + D^iD^jM_9- \left(\hat{\square}M_9+ 2 M_9 \dot{\bar{K}} \right) h^{ij} + 2M_9 \dot{\bar{K}}^{ij}  \right) D_i \pi D_j \pi - m_9(\tau,\rho) \pi^2\bigg] \,. 
\end{align}
Without any  $\dot{\pi}\pi'$ mixing, stability of this operator alone simply requires negativity for the  coefficient of $D_i \pi D_j \pi $,
\begin{align}
\left(M_9 \hat{R}^{ij} + D^iD^jM_9- \left(\hat{\square}M_9+ 2 M_9 \dot{\bar{K}} \right) h^{ij} + 2M_9 \dot{\bar{K}}^{ij}  \right) < 0 \,.
\end{align}

\item \underline{$M_{11}(\tau,\rho)\bar{K}^{\nu}_{\;\; \mu}\delta K \delta K^{\mu}_{\;\; \nu}  +M_{12}(\tau,\rho) \bar{K}^{\mu}_{\;\; \nu} \delta K^{\nu}_{\;\; \rho} \delta K^{\rho}_{\;\; \mu} $}: A similar pattern emerges for these operators. Only the combination 
\begin{align}
 M_{11} \left( \bar{K}^{\mu}_{\;\; \nu}\delta K \delta K^{\nu}_{\;\; \mu}  -\bar{K}^{\mu}_{\;\; \nu} \delta K^{\nu}_{\;\; \rho} \delta K^{\rho}_{\;\; \mu}\right)
\end{align}   
generates second-order equations for $\pi$. The associated contribution to the action is then
\begin{align}
 &\int \rd^4x \sqrt{-g} M_{11}(r) \left( \bar{K}^{\mu}_{\;\; \nu}\delta K \delta K^{\nu}_{\;\; \mu}  -\bar{K}^{\mu}_{\;\; \nu} \delta K^{\nu}_{\;\; \rho} \delta K^{\rho}_{\;\; \mu} \right) \nonumber \\
 & ~~~\rightarrow  \int \rd^4x \sqrt{-g}\bigg[ \bigg( D_kD^i  \left(M_{11}\bar{K}^{k j} \right)- \frac{1}{2}D_k D_l \left(M_{11} \bar{K}^{kl} \right)h^{ij} -\frac{1}{2} D_k D^k \left(M_{11}\bar{K}^{ij} \right)+  M_{11}\bar{K}^{k\ell}\hat{R}^{\; \,i \;\, j}_{k\;\ell} \nonumber \\
 &\quad ~~~~~~~~~~~~~~~~~~~~~ + M_{11} \dot{\bar{K}} \bar{K}^{\ij} + M_{11} \bar{K}^{kl}\dot{\bar{K}}_{kl} h^{ij} - 2 M_{11}\bar{K}^{k i} \dot{\bar{K}}_{k}^{\;\; j} \bigg)D_i \pi D_j \pi - m_{11}(\tau,\rho) \pi^2  \bigg]  \,.
 \end{align}
 Stability condition is also straight forward as there is no $\dot{\pi}\pi'$ mixing.

\item \underline{$M_{13}(\tau,\rho) \delta g^{\tau\tau} \delta \hat{R}$}: This operator gives rise to the following terms in the~$\pi$ action:
\begin{align}
&\int \rd^4x \sqrt{-g} M_{13} \delta g^{\tau\tau} \delta \hat{R} \nonumber \\
& ~~~\rightarrow  \int \rd^4x \sqrt{-g}\bigg[ 4 \left( M_{13}\left(h^{k\ell} \dot{\hat{\Gamma}}^m_{\ell m}-h^{\ell m} \dot{\hat{\Gamma}}^k_{\ell m} \right) + D_i\left(M_{13}\bar{K}^{ik}\right) - D^k\left(M_{13}\bar{K}\right)  \right)\dot{\pi}  D_k\pi \nonumber \\
 &\quad ~~~~~~~~~~~~~~~~~~~~~  + 2 \partial_{\tau}\left( M_{13}\bar{K} h^{ij}- M_{13} \bar{K}^{ij} \right) D_i\pi D_j\pi  - m_{13}(\tau,\rho) \pi^2 \bigg] \,,
\end{align}
where we have used the fact that~$\dot{\bar{h}}_{ij} = 2 \bar{K}_{ij}$. Again, it is not hard to see that there is no mixing of the form~$\dot{\pi} \partial_A \pi $. It can be diagonalized in the usual way but the resulting expression is lengthy and not informative which we do not display here. The stability constraints arising from these operators are rather less informative. We will restrict in Appendix~\ref{app:SdSlimit} to de Sitter-Schwarszchild limit to extract explicit constraints. 

\item  \underline{$M_{14}(\tau,\rho) \bar{K}^{\mu}_{\;\;\nu} \delta g^{\tau\tau} \delta \hat{R}^{\nu}_{\;\; \mu}$}: The quadratic $\pi$ action generated from this operator is
\begin{align}
&   \int \rd^4x \sqrt{-g} M_{14}\bar{K}^{\mu}_{\;\; \nu} \delta g^{\tau\tau} \delta \hat{R}^{\nu}_{\;\; \mu} \nonumber \\
& ~~~\rightarrow  \int \rd^4x \sqrt{-g} \bigg[ 2 \bigg( M_{14}\left( \dot{\hat{\Gamma}}^m_{km}\bar{K}^{ik} + \dot{\hat{\Gamma}}^m_{\ell k}\bar{K}^{\ell}_{\;\,m} h^{ki} -  \dot{\hat{\Gamma}}^m_{k\ell} h^{k\ell} \bar{K}_{m}^{\;\; i} - \dot{\hat{\Gamma}}^i_{k\ell }\bar{K}^{k\ell} \right) \nonumber \\
&\quad~~~~~~~~~~~~~~~~~~~~~ + D_k \left(2M_{14} \bar{K}^{i\ell} \bar{K}_\ell^{\;\,k} -M_{14}\bar{K}K^{ik}\right)  - D^i \left(M_{14} \bar{K}^{k\ell}\bar{K}_{k\ell} \right) \bigg) \dot{\pi} D_i\pi  \nonumber \\  
& \quad~~~~~~~~~~~~~~~~~~~~~ + \partial_{\tau}\left( M_{14} \bar{K}^{k\ell}\bar{K}_{k\ell}h^{ij} + M_{14}\bar{K} \bar{K}^{ij} - 2 M_{14}\bar{K}^{i\ell} \bar{K}_{\ell}^{\;\, j}\right) D_i\pi D_j\pi  - m_{14}(\tau,\rho)\pi^2 \bigg] \,.
\end{align}
After evaluating the form of $\hat{\Gamma}$'s,  one finds that there is only $\dot{\pi} \pi'$ mixing.

\end{itemize}
In principle, one should consider an action with all these terms together in order to come up with a constraint. However, such expression is not particularly illuminating here with many free functions. We stress again that instability of each term signals a potential instability in the entire theory although identifying the root cause of the instability in the Horndeski or DHOT theory may be tricky as the different operators considered here are likely not independent in the underlying theory. In Appendix~\ref{app:SdSlimit}, we collect the above constraints in the Schwarzschild-de Sitter limit.

\section{Odd-parity perturbations}
In this Section we analyze the effective action~\eqref{eqn:EFT} for odd-parity perturbations for~$\ell \ge 2$. The perturbed metric~$\delta g^{\rm odd}_{\mu\nu}$ is expanded in terms of the standard odd-parity vector and tensor spherical harmonics \cite{Regge:1957td} as
\begin{align}
\delta g^{\rm odd}_{\mu\nu}  = \sum_{\ell m}\begin{pmatrix}
0 & 0 &  h_0^{\ell m} \epsilon_{A}^{\;\;C} \nabla_C  \\
0 & 0 &  h_1^{\ell m} \epsilon_{A}^{\;\;C} \nabla_C  \\
h_0^{\ell m} \epsilon_{B}^{\;\;C} \nabla_C  &h_1^{\ell m} \epsilon_{B}^{\;\;C} \nabla_C & h_2^{\ell m}  \epsilon_{(B}^{\quad C} \nabla_{A)}\nabla_C 
\end{pmatrix} Y_{\ell}^m(\theta ,\varphi) \,,
\end{align} 
where~$\nabla_A$ is the covariant derivative on the 2-sphere associated with~$\gamma_{AB}$. As usual, we work in the Regge-Wheeler gauge 
\begin{align}
h_2^{\ell m}=0\, ,
\end{align}  
which is justified since the EFT is invariant under diffeomorphisms on the 2-sphere. Note that $h_2$ is exactly zero for $\ell =1$.

Odd perturbations on the 2-sphere are only affected by a restricted set of operators, since odd-parity contributions to~$\delta g^{\tau \tau}$ and~$\delta K$ are non-zero only starting at second order. For the same reason, odd sector perturbations correspond only to tensor perturbations. Therefore, at linear level the relevant EFT is 
\begin{align} \label{eqn:oddEFT}
 S_{(2)} = \int \rd ^4x \sqrt{-g}\bigg[& \frac{M_{\rm Pl}^2}{2}R - \Lambda(\tau,\rho) + \alpha(\tau,\rho) g^{\tau\tau}   + \beta(\tau,\rho) \bar{K}^{\mu}_{\;\; \nu}K^{\nu}_{\;\;\mu} \nonumber \\
 & + M_{10}(\tau,\rho) \delta K^{\mu}_{\;\; \nu}\delta K^{\nu}_{\;\; \mu}  + M_{12}(\tau,\rho)\bar{K}^{\mu}_{\;\; \nu} \delta K^{\nu}_{\;\,\rho} \delta K^{\rho}_{\;\; \mu} \bigg]\,,
\end{align}
where once again we have chosen to work in Einstein frame, in which~$M_1(\tau, \rho) = M_{\rm Pl}^2$. Furthermore note once more that we are leaving the coefficient functions as general functions of~$(\tau,\rho)$, specializing to functions of~$r(\tau,\rho)$ only when the analysis is restricted to Schwarzschild-de Sitter space-time.

In general, the action for~$h_0$ and~$h_1$ for a specific value of~$(\ell,m)$ takes the form
\begin{align}
 S = \int \rd \tau \rd \rho \left[ c \left( h'_0 - \dot{h}_1 + a h_0 + b h_1 \right)^2 + \frac{1}{2}k_{00} h_0^2 + k_{01}h_0 h_1 + \frac{1}{2} k_{11} h_1^2  \right]\,,
\end{align}
where all the coefficient functions $c$, $a$, $b$, $k_{00}$, $k_{01}$ and $k_{11}$ are functions of $(\tau, \,\rho)$. Restricting to the EFT~\eqref{eqn:oddEFT}, these are then explicitly given by
\begin{align} \label{eqn:oddcoeff}
  c &= \frac{M_{\rm Pl}^2}{4 \sqrt{F}}+\frac{M_{10}}{2 \sqrt{F}}+\frac{M_{12} \left(r^2 F\right)\dot{} }{8 F^{3/2} r^2} \,;\nonumber \\
  a &= -2 \frac{r'}{r}\,;\nonumber \\
b&=\frac{\dot{F}}{2 F}+\frac{\dot{r}}{r} -r^2 \frac{\beta +M_{\rm Pl}^2 }{ 8 c F^{3/2}} \left(\frac{F}{r^2}\right)^{\centerdot}\,;\nonumber \\
k_{00} &=\sqrt{F} (\alpha -\Lambda )+\frac{M_{\rm Pl}^2 \left(F^2 \left(j^2+2 \dot{r}^2\right)-2 F \left(r \left(2 r''-\dot{F} \dot{r}\right)+r'^2\right)+2 r F' r'\right)}{2F^{3/2} r^2} \nonumber \\
&\quad\, + M_{10} \frac{\sqrt{F} \left(j^2-2\right)}{r^2} +M_{12}\frac{\sqrt{F}  \dot{r} \left(j^2-2\right)}{r^3} \,;\nonumber \\
k_{01}& = \frac{M_{\rm Pl}^2 \left(2 F \dot{r}'- \dot{F} r'\right)}{F^{3/2} r}+\frac{r^2 \left(\dot{F}\left(\beta  F'-F \beta'\right)-\beta  F \dot{F}'\right)+4 \beta  F^2 r' \dot{r}-2 \beta  F r \dot{F}r'}{2 F^{5/2} r^2}\,;\nonumber \\
k_{11} &=M_{\rm Pl}^2 \frac{\left(-2 F^2 \left(j^2+4 r \ddot{r }\right)+r^2 \dot{F}^2+4 F r'^2-4 rF  \dot{F } \dot{r }\right)}{4 F^{5/2} r^2} \nonumber \\
&\quad \, +\frac{4 F^2 r^2 (\alpha +\lambda )+\beta  \left(r^2 \left(2 F \ddot{F }-\dot{F }^2\right)-4 F^2 \dot{r }^2\right)+2r^2 \dot{\beta  } F  \dot{F }}{4 F^{5/2} r^2}\nonumber \\
&\quad  \,
-\left(\beta +M_{\rm Pl}^2\right)^2\frac{\left(r \dot{F}-2 F \dot{r}\right)^2 }{32 c F^3 r^2}-\frac{c \left(r \dot{F}-2 F \dot{r }\right)^2}{2 F^2 r^2}\,,
\end{align}
where $F(\tau,\rho) = 1- f(r)$ and $j^2 = \ell^2+\ell$. The above expressions apply to a general case where $\Lambda,\, \alpha,\, \beta, \, M_{10},\, M_{12} $  are general functions of $(\tau,\, \rho)$. 

One can identify the relevant degrees of freedom by integrating in an auxiliary field~$\Psi$ as follows:
\begin{align}
  S= \int \rd \tau \rd \rho & \bigg[ c \left( h'_0 - \dot{h}_1 + a h_0 + b h_1 \right)^2
  - c\left( \frac{1}{c}\Psi - (h'_0 - \dot{h}_1 + a h_0 + b h_1) \right)^2 \nonumber \\
   & ~~~ + \frac{1}{2}k_{00} h_0^2 + k_{01}h_0 h_1 + \frac{1}{2} k_{11} h_1^2  \bigg] \,.
\end{align}
The equations of motion for $h_0$ and $h_1$ become,
\begin{align}
 \begin{pmatrix}
   k_{00} & k_{01} \\
   k_{01} & k_{11}
 \end{pmatrix} \begin{pmatrix}
  h_0 \\ h_1 
 \end{pmatrix} = \begin{pmatrix}
   2 \Psi' - 2a \Psi  \\
     -2 \dot{\Psi} - 2b \Psi 
 \end{pmatrix},
\end{align}
which can then be expressed purely in terms of $\Psi$ in the action,
\begin{align}
 S & = \int \rd \tau \rd \rho  \Bigg[ -\frac{1}{2} \begin{pmatrix}
  2 \Psi' - 2a \Psi  \\    -2 \dot{\Psi} - 2b \Psi 
 \end{pmatrix}^{\rm T}  \begin{pmatrix}
   k_{00} & k_{01} \\
   k_{01} & k_{11}
 \end{pmatrix}^{-1} \begin{pmatrix}
  2 \Psi' - 2a \Psi  \\    -2 \dot{\Psi} - 2b \Psi 
 \end{pmatrix} - \frac{1}{c} \Psi^2 \Bigg] \nonumber \\
 & = \int \rd \tau \rd \rho \Bigg[ \frac{2}{k_{01}^2 -k_{00} k_{11}} \left(k_{00} \dot{\Psi}^2 + 2k_{01} \dot{\Psi} \Psi' + k_{11} \Psi'^2 \right) - \frac{1}{2} m_{\Psi} \Psi^2  \Bigg]  \,.
\end{align}
Stability of the Hamiltonian requires that 
\begin{align}
  k_{00} > 0, \quad  k_{11} < 0\,.
\end{align}
The kinetic terms can be diagonalized as usual, with the corresponding eigenvalues given by
\begin{align}
\frac{1}{2}\left( k_{00}+ k_{11} \pm (k_{00} -k_{11}) \left( 1+ \frac{4 k_{01}^2}{(k_{00} -k_{11})^2}\right)^{\frac{1}{2}}\right).
\end{align}
However, the resulting stability constraints on the functions $\alpha$, $\beta$, $M_{10}$ and  $M_{12}$ are unclear in general, due to the complicated forms of the coefficient functions~\eqref{eqn:oddcoeff}. 
 
\subsection{Schwarzschild-de Sitter limit}

We now restrict the analysis to the Schwarzschild-de Sitter case, in order to obtain useful constraints. As mentioned in Sec.~\ref{sec:isometry}, the background isometries constrain the EFT coefficients to be functions of~$r(\tau,\rho)$, 
\begin{align}
\Lambda(\tau,\rho) & \to \Lambda(r)\,;  \nonumber \\
\alpha(\tau,\rho) &\to \alpha(r)\,;  \nonumber \\
\beta(\tau,\rho) &\to \beta(r)\,;  \nonumber \\
M_{10}(\tau,\rho) &\to M_{10}(r)\,;  \nonumber \\
M_{12}(\tau,\rho) &\to M_{12}(r)\,. 
\end{align}
One can show that, under the condition $g(r) = f(r)$, together with the background equations of motion~\eqref{eqn:background}, 
\begin{align}
 k_{01} =0\,.
\end{align}
This holds even without using the explicit form of~$f(r)$. Therefore, stability requires that 
\begin{align} \label{eqn:kconstraints}
k_{00} > 0\,;\qquad 
k_{11}  < 0 \,,
\end{align}
with
\begin{align}
k_{00} & = \frac{\sqrt{1-f} \left(j^2-2\right) \left(r M_{\rm Pl}^2+2rM_{10} -2\sqrt{1-f} M_{12}\right)}{r^3}\,; \nonumber \\   
 k_{11}  & =  \frac{\left(2 (1-f) +rf_r\right){}^2 \left(2 r\sqrt{1-f}  \left(\beta -2 M_{10}\right)+M_{12} \left(2 (1-f) - r f_r\right)\right){}^2}{16r^3(1-f)^2\left(M_{12} \left(2 (1-f)-r f_r\right)-2 (M_{\rm Pl}^2+2M_{10})r \sqrt{1-f}   \right)} \,, \nonumber \\   
 &\quad \, -M_{\rm Pl}^2 \frac{\left(j^2-2\right) }{2r^2 \sqrt{1-f} }
\end{align}
where we have defined~$j^2 \equiv \ell (\ell + 1)$. In the above we have used the background equations to eliminate~$\Lambda$,~$\alpha$ and~$\beta_r$
in terms of~$f$ and~$\beta$. Furthermore, one finds that $\alpha=\beta=0$ and $\Lambda= 2 M_{\rm Pl}^2 \lambda$ in the Schwarzschild-de Sitter limit, with $f(r) =g(r) = 1-\frac{r_s}{r} - \frac{\lambda}{3} r^2$.  Therefore, the radial sound speed for perturbations in the Schwarzschild-de Sitter limit is 
\begin{align} \label{eqn:cssq}
 c_{\rho}^2 &= -\frac{g^{\tau\tau}k_{11} }{  g^{\rho\rho} k_{00}} \nonumber \\
 &=  \frac{1}{8 (1-f)^{3/2} \left(j^2-2\right) \left(2 \sqrt{1-f} M_{12}-r M_{\rm Pl}^2-2r M_{10} \right) \left(M_{12} \left(r f_r+2 f-2\right)+2 r \sqrt{1-f}  \left(M_{\rm Pl}^2+2 M_{10}\right)\right)}\nonumber\\ 
   & \qquad  \times \bigg[  8   M_{\rm Pl}^2 r (1-f)^{3/2} \left(j^2-2\right)\left(4r M_{10} \sqrt{1-f}  +M_{12} \left(r f_r+2 f-2\right)\right) +16r^2 M_p^4 \left(j^2-2\right) (1-f)^2 \nonumber \\
   &\quad \qquad +4 M_{12}\left(\beta -2 M_{10}\right) r\sqrt{1-f}  \left(2-2 f+rf_r \right)^2 \left(2-2 f-r f_r\right) + M_{12}^2 \left( r^2f_1^2 -4 (1-f)^2\right)^2 \nonumber \\
     &\quad \qquad +4 \left(\beta -2 M_{10}\right)^2 r^2(1-f) \left(2-2 f+rf_r\right){}^2  \bigg]\, . 
\end{align}

The dependence on~$j^2=\ell(\ell+1)$ of the sound speed is due to the unknown coefficients~$M_{10}$ and~$M_{12}$. In a Horndeski theory with~${\cal L}_2$ and~${\cal L}_4$ only,\footnote{We are using the convention that 
\begin{align}
{\cal L}_2 &= P(X), \nonumber \\
{\cal L}_4 &=G_4(X) R + G_{4,X}(X)\big( (\square \phi)^2 - \phi_{\mu\nu}\phi^{\mu\nu} \big),
\end{align}
where $X= - \frac{1}{2} \partial_{\mu}\phi \partial^{\mu} \phi$.
} 
$P(X)$, $G_4(X)$ and their derivatives evaluated on the background have to satisfy the following relations on the  Schwarzschild-de Sitter geometry,
\begin{align}
  \bar{P} +2 \lambda \big(\bar{G}_4 - m^4\bar{G}_{4,X}\big) =0 \,; \nonumber \\
    \bar{P}_{,X} + 2 \lambda \big(\bar{G}_{4,X} + m^4\bar{G}_{4,XX}\big) =0 \,.
\end{align}
After using these relations, one finds that~$c_s^2 $ coming from Horndeski ${\cal L}_2$ and ${\cal L}_4$ is simply
\begin{align}
 c_s^2 = \frac{\bar{G}_4}{\bar{G}_4 - m^4\bar{G}_{4,X}}.
\end{align}
Note that a quadratic DHOST gives the same sound speed in the odd sector~\cite{Takahashi:2021bml}. The $\ell$ dependence in \eqref{eqn:cssq} will cancel out and give the exact same expression as in Horndeski if one chooses $M_{12}=0$ and $M_{10}$ being the corresponding coefficient from Horndeski~${\cal L}_4$. Phenomenologically speaking, apart from the stability constraints \eqref{eqn:kconstraints}, there is also stringent observational restrictions on the propagation speed of gravitational waves from neutron star mergers~\cite{TheLIGOScientific:2017qsa,GBM:2017lvd,Monitor:2017mdv}. In such cases, the relevant gravitational wave perturbation is the lowest $\ell =2$ mode, given the current experimental signal-to-noise ratio.
 
\section{Discussion}
Effective field theory is a powerful tool for the study of perturbative systems or of the macroscopic behavior of theories at low energies, which is a typical situation that arises in many different areas of physics. The robustness of EFT allows us to investigate a large class of theories, in our case modified gravity theories, using a unified framework. Furthermore, it has a wide range of applicability in extreme limits of gravitational systems, such as in cosmological inflation and in black holes. One particularly interesting application has been to scalar-tensor theories which, over the past several decades, have drawn significant attention due to their rich phenomenological applications in both early and late-time cosmology. While one well-studied application has been to the physics of inflation, the technique can also be used to study other interesting approaches to the early universe, including those in which a time-dependent scalar profile drives cosmological expansion through a conformal coupling with matter, {\it e.g.},~\cite{Rubakov:2009np,Creminelli:2010ba,Hinterbichler:2011qk}. Scalar profiles of this type provide an interesting connection between cosmological evolution and astrophysical black holes, which can emerge as allowed non-trivial solutions with time-like hair is generic in scalar tensor theories. 

In this work we have investigated the EFT of perturbations around such black holes with time-like scalar hair. Building on the construction in \cite{Mukohyama:2005rw}, we argue that the EFT coefficients must satisfy certain constraints imposed by the background isometries. When the underlying scalar-tensor theory enjoys a shift symmetry in~$\phi$, the combination between this symmetry and time-translation invariance (in Schwarszchild type coordinates) remains as an unbroken symmetry, and constrains the EFT coefficient functions, despite the fact that the time-like scalar profile spontaneously breaks both symmetries individually. We have constructed a general set of operators, up to second-order in derivatives, which are compatible with the aforementioned symmetries.  Motivated by the fact that gradient instabilities appear in the scalar sector in both the Horndeski and DHOST theories~\cite{Khoury:2020aya,Takahashi:2021bml}, we have performed a stability analysis in the decoupling limit, which we argue should capture the essential properties of scalar perturbations. In addition, we have shown that odd sector perturbations can be comprehensively analyzed despite the appearance of many unknown functions, since only a few terms in the action contribute to odd-parity perturbations. This analysis also provides hints as to how we might construct potentially stable black hole/wormhole solutions. 

In order to complete the story, a full analysis of even-parity perturbations is required. Such a task promises to be technically challenging, despite the fact that a subset of the EFT operators (essentially those corresponding to the quadratic DHOST terms, for instance $\delta K^2$) can be analyzed using the technique in~\cite{Takahashi:2021bml}.  However, Inclusion of operators of the type $\bar{K}_{\mu\nu}\delta K^{\rho \mu} \delta K^{\nu}_{\;\; \rho}$ would require different techniques in order to identify the correct propagating degrees of freedom, and to perform the stability analysis. 

It is also worth commenting on higher-derivative terms in the EFT. The inclusion of higher-derivative operators does not necessarily mean higher-order equations of motion for perturbations. In fact, we expect that a similar story to the DHOST analysis should play out for perturbations, so that a degeneracy condition between coefficient functions of higher-derivative operators can be imposed, reducing the equation of motion to second order. Furthermore, it is possible for higher derivative operators to contribute to a non-vanishing dispersion relation~\cite{ArkaniHamed:2003uy,Babichev:2018twg}, or to serve as ``Scordatura terms" \cite{Motohashi:2019ymr,Gorji:2020bfl,DeFelice:2022xvq}, in such a way as to solve the strong coupling problem for some solutions. 

In future work, we intend to extend this work to the case of rotating black holes, which comprise the majority of astrophysical black holes observed through current and future instruments. The EFT for slowly rotating black holes with time-independent scalar hair has already been constructed~\cite{Hui:2021cpm}, while interesting background solutions for rotating black holes with time-like hair have found in DHOST models~\cite{Charmousis:2019vnf,BenAchour:2020fgy}. The EFT description of perturbations around these objects remains to be constructed, and may be complicated by the fact that the background construction relies on the use of disformal transformations. It is an interesting future exercise to investigate whether similar techniques to those used in this paper can be applied to perturbations and the construction of the EFT in these rotating backgrounds.

\bigskip
While this paper was in preparation, we were aware of the research by Mukohyama, Takahashi and Yingcharoenrat on a very similar subject \cite{Mukohyama:2022skk}. Our results are consistent with each other on the analysis of odd parity sector perturbations. We thank them for their kind correspondence. 

\bigskip
\goodbreak
\centerline{\bf Acknowledgements}
We thank Hayato Motohashi, Luca Santoni, Kazufumi Takahashi and Enrico Trincherini  for useful discussions. This work is supported in part by the US Department of Energy (HEP) Award DE-SC0013528, NASA ATP grant 80NSSC18K0694, and by the Simons Foundation Origins of the Universe Initiative. T.N. is supported in part by JSPS KAKENHI Grant No.~20H01902 and No.~22H01220, and MEXT KAKENHI Grant No.~21H00075, No.~21H05184 and No.~21H05462.
\appendix
\section{Background equations of motion} \label{app:bgEOM}

The tadpole terms of~\eqref{eqn:EFT},
 \begin{align}
S_{\rm tadpole} =   \int\rd^4x \sqrt{-g}\Big[ \frac{M_1}{2} R - \Lambda + \alpha g^{\tau\tau} + \beta \bar{K}_{\;\;\mu}^{\nu} K^{\mu}_{\;\;\nu} \Big]\,,
 \end{align}
generate the background equation of motion,
\begin{align}
\left( \bar{R}_{\mu\nu} - \frac{1}{2} \bar{g}_{\mu\nu} \bar{R}  - \bar{\nabla}_{\mu}\bar{\nabla}_{\nu} + \bar{g}_{\mu\nu} \bar{\square} \right)M_1 = \bar{T}_{\mu\nu}\,,
\end{align}
where
\begin{align}
 \bar{T}_{\mu\nu} &= \Big( \alpha g^{\tau\tau} - \Lambda  + \beta \bar{K}_{\;\;\rho}^{ \sigma}\bar{K}_{\;\;\sigma}^{ \rho} \Big) \bar{g}_{\mu\nu} - 2 \alpha \delta_{\mu}^{\tau} \delta_{\nu}^{\tau}  - 2 \beta \bar{K}^{\rho}_{\;\;\mu} \bar{K}_{\rho \nu} + \beta  \bar{K}^{\sigma}_{\;\; \rho}\bar{K}^{ \rho}_{\;\; \sigma}  n_{\mu} n_{\nu}  \nonumber \\
 &\qquad + \bar{\nabla}_{\lambda} \Big( \beta \bar{K}_{\;\; \mu}^{\lambda} n_{\nu} + \beta \bar{K}_{\;\; \nu} ^{\lambda} n_{\mu} - \beta \bar{K}_{\mu\nu} n^{\lambda}  \Big) \,.
\end{align}

Here we collect the useful equations of motion evaluated on the ansatz~\eqref{eqn:glemaitre}, with $r$-dependent coefficient functions and constant $M_1 = M_{\rm Pl}^2$,
\begin{align}\label{eqn:background}
\frac{\delta}{\delta g^{\tau\tau} } : & \quad 0=r^2 (\alpha -\Lambda )+ M_1\frac{ r (1-f) g f_r +f^2 (1-g)+ r f g_r }{f^2} \,; \nonumber \\
\frac{\delta}{\delta g^{\tau\rho} } : &  \quad 0= 2 M_1 r\frac{ (f-1)  \left(g f_r-f g_r\right)}{f}+g r^2 f_r \beta _r \nonumber \\
&\qquad \quad +\beta  \left(\frac{1}{2} r^2 \left(\frac{(2 f-1) g f_r^2}{(1-f) f}+f_r g_r+2 g f_{rr}\right)+2 g r f_r+4 (1-f) g\right)\,;\nonumber \\
\frac{\delta}{\delta g^{\rho\rho} } : &  \quad 0= \frac{M_1}{2} \left(\frac{4 r \left((1-f) f g_r-g f_r \right)}{f}+ 4 f (1-g)\right)+r^2 \left(g f_r \beta _r-2 f (\alpha +\Lambda )\right)\nonumber \\
&\qquad \quad +\beta  \left(\frac{1}{2} r^2 \left(\frac{(1-2f)g f_r^2}{(f-1) f}+f_r g_r+2 g f_{rr}\right)+2r  g  f_r + 4 (1-f) g\right)\,; \nonumber \\
\frac{\delta}{\delta g^{\theta\theta} }  \mbox{ or }  \frac{\delta}{\delta g^{\phi\phi} } : &  \quad 0=  M_1 \left(  r^2\frac{ f f_r g_r- g f_r^2 +2 f g f_{rr}  }{f}+2 r (f_r g+ fg_r)\right) +4 r^2f (\alpha +\Lambda ) \nonumber \\
&\qquad \quad +\beta  \left(2 r \frac{(1-f) f g_r-(1+f) g f_r }{f}+\frac{  r^2 g f_r^2}{(f-1) } - 4 g(1-f)\right)+4 r \beta_r (1-f) g  \,.
\end{align}
These are used in our analysis to re-express~$\Lambda$,~$\alpha$,~$\beta$ and~$\beta_r$ in terms of $f(r)$, $g(r)$ and their derivatives.

\section{The St\"uckelberg procedure}
This Appendix contains the complete expressions for the appearance of the St\"uckelberg field $\pi$.
\begin{align}
 g^{\tau\tau} &\to g^{\tau\tau}(1 + 2\dot{\pi} + \dot{\pi}^2) + 2 g^{a \tau} (\partial_{a}\pi +  \dot{\pi} \partial_a \pi)  + g^{ab} \partial_a \pi \partial_b \pi, \nonumber\\
 g^{ a \tau} & \to g^{a \tau  } (1+ \dot{\pi})  + g^{ab}\partial_b\pi, \nonumber \\
 g^{ab} &\to g^{ab}.
\end{align}
The derivatives $\partial_{\mu}$ transform as 
\begin{align}
   \partial_{\tau} &\to (1- \dot{\pi} + \dot{\pi}^2 ) \partial_{\tau} + {\cal O}(\pi^3)\,; \nonumber \\
   \partial_i &\to \partial_i - (1-\dot{\pi})\partial_i \pi \partial_{\tau}+ {\cal O}(\pi^3)\,.
\end{align}
To calculate the transformation of~$\hat{R}_{ij}$ in~\eqref{eqn:curvpi}, we have used
\begin{align}
\hat{R}_{ij} &\to \hat{R}_{ij} - \partial_k \pi \dot{\hat{\Gamma}}^k_{ij} +\partial_i \pi \dot{\hat{\Gamma}}^k_{jk} + D_k \left( \delta \hat{\Gamma}^k_{ij} \right) - D_i \left( \delta \hat{\Gamma}^k_{kj} \right)  \nonumber \\
D_k \dot{h}_{li} & =\dot{\hat{\Gamma}}^m_{li} h_{mi} + \dot{\hat{\Gamma}}^m_{ki} h_{lm} \,.
\end{align}

\section{Decoupling limit in Schwarszchild-de Sitter} \label{app:SdSlimit}
It is convenient to collect the results in section \eqref{sec:pistability} restricted to the Schwarszchild-de Sitter geometry. In this case the metric is given by~\eqref{eqn:gtr} with~$f(r) =g(r) = 1-\frac{r_s}{r} - \frac{\lambda}{3} r^2$, where~$0<\lambda < \frac{4}{9 r_s^2}$ for two horizons~$r_-,r_+$ to exist, and the inner horizon~$r_-$ satisfies~$1<\frac{r_-}{r_s}<\frac{3}{2}$. The individual kinetic terms  for $\pi$ are:

\begin{itemize}

\item \underline{$M_3(\tau,\rho)\, \delta g^{\tau\tau} \delta K$}: 
\begin{align}
\frac{2M_{3,r}}{\sqrt{F}}\dot{\pi}\pi'+ \frac{Fr^2 M_{3,r} +M_3 (3 r_s-2 rF )}{r^2F^{3/2} }\pi'^2 + \frac{\sqrt{F} \left(r M_{3,r} -2 M_3\right)}{r^3}\gamma^{AB}\partial_A\pi \partial_B\pi
\end{align}
where $F(r)= \frac{r_s}{r}+\frac{\lambda}{3}r^2 $, and $\left(\frac{3}{2}\right)^{2/3} r_s^{2/3}\lambda^{1/3} <F<1$ between two horizons. 
\item \underline{$M_4(\tau,\rho) \bar{K}_{\mu\nu}\delta g^{\tau\tau} \delta K^{\mu\nu}$}:
 \begin{align}
&\quad \frac{2 F  r^2 (3 r_s-2 F r) M_{4,r}+9 r_s^2 M_4}{2 F^2 r^4}\dot{\pi}\pi'+\frac{M_4\left(8 F^2 r^2-36 F r r_s+27 r_s^2\right)+2 F  r^2 (3 r_s-2 F r) M_{4,r}}{4 F^2 r^4}\pi'^2\nonumber \\
&+ \frac{(4 F r +3 r_s )M_4-2 F  r^2 M_{4,r}}{2 r^5}\gamma^{AB}\partial_A\pi \partial_B\pi
\end{align}

\item \underline{$M_9 \left(\delta K^2 - \delta K_{\mu\nu} \delta K^{\mu\nu}\right) $}: 
\begin{align}
&\quad \frac{M_{9} \left(-4 F^2 r^2+18 F r r_s-9 r_s^2\right)-2 Fr^3 M_{9,r} }{F^2 r^4}\pi'^2 \nonumber\\
&-\left( \frac{M_{9} (4 F r-3 r_s) (2 F r+3 r_s)}{2 F r^6}+\frac{r M_{9,rr} +M_{9,r}}{r^3} \right)\gamma^{AB}\partial_A\pi \partial_B\pi
\end{align}

\item \underline{$ M_{11} \left( \bar{K}_{\mu\nu}\delta K \delta K^{\mu\nu}  -\bar{K}_{\mu\nu}\delta K^{\rho \mu} \delta K^{\nu}_{\;\; \rho} \right) $}:
\begin{align}
&\quad \frac{2 F r^3 (4 F r-3 r_s) M_{11,r} -3 M_{11} \left(8 F^3 r^3-20 F^2 r^2 r_s+3 (8 F+1) r r_s^2-9 r_s^3\right)}{4 F^{5/2} r^6}\pi'^2\nonumber \\
&+\frac{1}{16 F^{5/2} r^8} \Big[4 F r^2 \left( \left(4 F^2 r^2-9 r_s^2\right)M_{11,r}  +F r^2 (4 F r-3 r_s) M_{11,rr} \right) \nonumber \\ &+M_{11} \left(-96 F^4 r^3+96 F^3 r^2 r_s+36 (3-2 F) F r r_s^2+27 (2 F-3) r_s^3\right) \Big] \gamma^{AB}\partial_A\pi \partial_B\pi
\end{align}

\item \underline{$M_{13}(\tau,\rho) \delta g^{\tau\tau} \delta \hat{R}$}:
\begin{align}
&\quad \frac{8M_{13,r}}{r}\dot{\pi}\pi'+\left[\frac{M_{13} (6 r_s-8 F r)}{F r^3}+\frac{4 M_{13,r}}{r} \right]\pi'^2\nonumber \\
&+ \left[\frac{(4 F r-3 r_s)}{r^4}M_{13,r} +M_{13} \left(-\frac{9 r_s^2}{2 F r^6}-\frac{8 F}{r^4}+\frac{9 r_s}{r^5}\right)\right]\gamma^{AB}\partial_A\pi \partial_B\pi
\end{align}

\item  \underline{$M_{14}(\tau,\rho) \bar{K}_{\mu\nu} \delta g^{\tau\tau} \delta \hat{R}^{\mu\nu}$}: 
\begin{align}
&\quad  \frac{9 r_s^2 \left(2 M_{14}+r M_{14,r} \right)-4 F^2 r^3 M_{14,r}}{2 F^{3/2} r^5} \dot{\pi} \pi' \nonumber \\
&+\frac{M_{14} \left(8 F^2 r^2-15 F r r_s+9 r_s^2\right)+F r^2 (3 r_s-4 F r)M_{14,r}  }{F^{3/2} r^5}\pi'^2 \nonumber \\
&+ \frac{F r^2 \left(-16 F^2 r^2+18 F r r_s-9 r_s^2\right) M_{14,r} +M_{14} \left(32 F^3 r^3-42 F^2 r^2 r_s+72 F r r_s^2-27 r_s^3\right)}{4 F^{3/2} r^8}\gamma^{AB}\partial_A\pi \partial_B\pi
\end{align}

\end{itemize}

\renewcommand{\em}{}
\bibliographystyle{utphys}
\addcontentsline{toc}{section}{References}
\bibliography{BH_hair}

\providecommand{\href}[2]{#2}\begingroup\raggedright\begin{thebibliography}{10}

\bibitem{TheLIGOScientific:2016pea}
{\bf LIGO Scientific, Virgo} , B.~Abbott {\em et al.}, ``{Binary Black Hole
  Mergers in the first Advanced LIGO Observing Run},''
  \href{http://dx.doi.org/10.1103/PhysRevX.6.041015}{{\em Phys.\ Rev.\ X} {\bf
  6} (2016) no.~4, 041015}, \href{http://arxiv.org/abs/1606.04856}{{\tt
  arXiv:1606.04856 [gr-qc]}}. [Erratum: Phys.Rev.X 8, 039903 (2018)].

\bibitem{Akiyama:2019cqa}
{\bf Event Horizon Telescope} , K.~Akiyama {\em et al.}, ``{First M87 Event
  Horizon Telescope Results. I. The Shadow of the Supermassive Black Hole},''
  \href{http://dx.doi.org/10.3847/2041-8213/ab0ec7}{{\em Astrophys.\ J.} {\bf
  875} (2019) no.~1, L1}, \href{http://arxiv.org/abs/1906.11238}{{\tt
  arXiv:1906.11238 [astro-ph.GA]}}.

\bibitem{Regge:1957td}
T.~Regge and J.~A. Wheeler, ``{Stability of a Schwarzschild singularity},''
  \href{http://dx.doi.org/10.1103/PhysRev.108.1063}{{\em Phys. Rev.} {\bf 108}
  (1957)  1063--1069}.

\bibitem{Zerilli:1970se}
F.~J. Zerilli, ``{Effective potential for even parity Regge-Wheeler
  gravitational perturbation equations},''
  \href{http://dx.doi.org/10.1103/PhysRevLett.24.737}{{\em Phys. Rev. Lett.}
  {\bf 24} (1970)  737--738}.

\bibitem{Berti:2009kk}
E.~Berti, V.~Cardoso, and A.~O. Starinets, ``{Quasinormal modes of black holes
  and black branes},''
  \href{http://dx.doi.org/10.1088/0264-9381/26/16/163001}{{\em Class.\ Quant.\
  Grav.} {\bf 26} (2009)  163001}, \href{http://arxiv.org/abs/0905.2975}{{\tt
  arXiv:0905.2975 [gr-qc]}}.

\bibitem{Woodard:2015zca}
R.~P. Woodard, ``{Ostrogradsky's theorem on Hamiltonian instability},''
  \href{http://dx.doi.org/10.4249/scholarpedia.32243}{{\em Scholarpedia} {\bf
  10} (2015) no.~8, 32243}, \href{http://arxiv.org/abs/1506.02210}{{\tt
  arXiv:1506.02210 [hep-th]}}.

\bibitem{Solomon:2017nlh}
A.~R. Solomon and M.~Trodden, ``{Higher-derivative operators and effective
  field theory for general scalar-tensor theories},''
  \href{http://dx.doi.org/10.1088/1475-7516/2018/02/031}{{\em JCAP} {\bf 02}
  (2018)  031}, \href{http://arxiv.org/abs/1709.09695}{{\tt arXiv:1709.09695
  [hep-th]}}.

\bibitem{Horndeski:1974wa}
G.~W. Horndeski, ``{Second-order scalar-tensor field equations in a
  four-dimensional space},'' \href{http://dx.doi.org/10.1007/BF01807638}{{\em
  Int.\ J.\ Theor.\ Phys.} {\bf 10} (1974)  363--384}.

\bibitem{Nicolis:2008in}
A.~Nicolis, R.~Rattazzi, and E.~Trincherini, ``{The Galileon as a local
  modification of gravity},''
  \href{http://dx.doi.org/10.1103/PhysRevD.79.064036}{{\em Phys. Rev. D} {\bf
  79} (2009)  064036}, \href{http://arxiv.org/abs/0811.2197}{{\tt
  arXiv:0811.2197 [hep-th]}}.

\bibitem{Deffayet:2009wt}
C.~Deffayet, G.~Esposito-Farese, and A.~Vikman, ``{Covariant Galileon},''
  \href{http://dx.doi.org/10.1103/PhysRevD.79.084003}{{\em Phys. Rev. D} {\bf
  79} (2009)  084003}, \href{http://arxiv.org/abs/0901.1314}{{\tt
  arXiv:0901.1314 [hep-th]}}.

\bibitem{Deffayet:2009mn}
C.~Deffayet, S.~Deser, and G.~Esposito-Farese, ``{Generalized Galileons: All
  scalar models whose curved background extensions maintain second-order field
  equations and stress-tensors},''
  \href{http://dx.doi.org/10.1103/PhysRevD.80.064015}{{\em Phys. Rev. D} {\bf
  80} (2009)  064015}, \href{http://arxiv.org/abs/0906.1967}{{\tt
  arXiv:0906.1967 [gr-qc]}}.

\bibitem{Deffayet:2011gz}
C.~Deffayet, X.~Gao, D.~Steer, and G.~Zahariade, ``{From k-essence to
  generalised Galileons},''
  \href{http://dx.doi.org/10.1103/PhysRevD.84.064039}{{\em Phys. Rev. D} {\bf
  84} (2011)  064039}, \href{http://arxiv.org/abs/1103.3260}{{\tt
  arXiv:1103.3260 [hep-th]}}.

\bibitem{Kobayashi:2011nu}
T.~Kobayashi, M.~Yamaguchi, and J.~Yokoyama, ``{Generalized G-inflation:
  Inflation with the most general second-order field equations},''
  \href{http://dx.doi.org/10.1143/PTP.126.511}{{\em Prog. Theor. Phys.} {\bf
  126} (2011)  511--529}, \href{http://arxiv.org/abs/1105.5723}{{\tt
  arXiv:1105.5723 [hep-th]}}.

\bibitem{Gleyzes:2014dya}
J.~Gleyzes, D.~Langlois, F.~Piazza, and F.~Vernizzi, ``{Healthy theories beyond
  Horndeski},'' \href{http://dx.doi.org/10.1103/PhysRevLett.114.211101}{{\em
  Phys. Rev. Lett.} {\bf 114} (2015) no.~21, 211101},
  \href{http://arxiv.org/abs/1404.6495}{{\tt arXiv:1404.6495 [hep-th]}}.

\bibitem{Langlois:2015cwa}
D.~Langlois and K.~Noui, ``{Degenerate higher derivative theories beyond
  Horndeski: evading the Ostrogradski instability},''
  \href{http://dx.doi.org/10.1088/1475-7516/2016/02/034}{{\em JCAP} {\bf 02}
  (2016)  034}, \href{http://arxiv.org/abs/1510.06930}{{\tt arXiv:1510.06930
  [gr-qc]}}.

\bibitem{Crisostomi:2016czh}
M.~Crisostomi, K.~Koyama, and G.~Tasinato, ``{Extended Scalar-Tensor Theories
  of Gravity},'' \href{http://dx.doi.org/10.1088/1475-7516/2016/04/044}{{\em
  JCAP} {\bf 04} (2016)  044}, \href{http://arxiv.org/abs/1602.03119}{{\tt
  arXiv:1602.03119 [hep-th]}}.

\bibitem{BenAchour:2016fzp}
J.~Ben~Achour, M.~Crisostomi, K.~Koyama, D.~Langlois, K.~Noui, and G.~Tasinato,
  ``{Degenerate higher order scalar-tensor theories beyond Horndeski up to
  cubic order},'' \href{http://dx.doi.org/10.1007/JHEP12(2016)100}{{\em JHEP}
  {\bf 12} (2016)  100}, \href{http://arxiv.org/abs/1608.08135}{{\tt
  arXiv:1608.08135 [hep-th]}}.

\bibitem{Takahashi:2017pje}
K.~Takahashi and T.~Kobayashi, ``{Extended mimetic gravity: Hamiltonian
  analysis and gradient instabilities},''
  \href{http://dx.doi.org/10.1088/1475-7516/2017/11/038}{{\em JCAP} {\bf 11}
  (2017)  038}, \href{http://arxiv.org/abs/1708.02951}{{\tt arXiv:1708.02951
  [gr-qc]}}.

\bibitem{Langlois:2018jdg}
D.~Langlois, M.~Mancarella, K.~Noui, and F.~Vernizzi, ``{Mimetic gravity as
  DHOST theories},''
  \href{http://dx.doi.org/10.1088/1475-7516/2019/02/036}{{\em JCAP} {\bf 02}
  (2019)  036}, \href{http://arxiv.org/abs/1802.03394}{{\tt arXiv:1802.03394
  [gr-qc]}}.

\bibitem{Kobayashi:2019hrl}
T.~Kobayashi, ``{Horndeski theory and beyond: a review},''
  \href{http://dx.doi.org/10.1088/1361-6633/ab2429}{{\em Rept. Prog. Phys.}
  {\bf 82} (2019) no.~8, 086901}, \href{http://arxiv.org/abs/1901.07183}{{\tt
  arXiv:1901.07183 [gr-qc]}}.

\bibitem{Sotiriou:2013qea}
T.~P. Sotiriou and S.-Y. Zhou, ``{Black hole hair in generalized scalar-tensor
  gravity},'' \href{http://dx.doi.org/10.1103/PhysRevLett.112.251102}{{\em
  Phys. Rev. Lett.} {\bf 112} (2014)  251102},
  \href{http://arxiv.org/abs/1312.3622}{{\tt arXiv:1312.3622 [gr-qc]}}.

\bibitem{Sotiriou:2014pfa}
T.~P. Sotiriou and S.-Y. Zhou, ``{Black hole hair in generalized scalar-tensor
  gravity: An explicit example},''
  \href{http://dx.doi.org/10.1103/PhysRevD.90.124063}{{\em Phys. Rev. D} {\bf
  90} (2014)  124063}, \href{http://arxiv.org/abs/1408.1698}{{\tt
  arXiv:1408.1698 [gr-qc]}}.

\bibitem{Babichev:2016rlq}
E.~Babichev, C.~Charmousis, and A.~Lehébel, ``{Black holes and stars in
  Horndeski theory},''
  \href{http://dx.doi.org/10.1088/0264-9381/33/15/154002}{{\em Class. Quant.
  Grav.} {\bf 33} (2016) no.~15, 154002},
  \href{http://arxiv.org/abs/1604.06402}{{\tt arXiv:1604.06402 [gr-qc]}}.

\bibitem{Benkel:2016rlz}
R.~Benkel, T.~P. Sotiriou, and H.~Witek, ``{Black hole hair formation in
  shift-symmetric generalised scalar-tensor gravity},''
  \href{http://dx.doi.org/10.1088/1361-6382/aa5ce7}{{\em Class. Quant. Grav.}
  {\bf 34} (2017) no.~6, 064001}, \href{http://arxiv.org/abs/1610.09168}{{\tt
  arXiv:1610.09168 [gr-qc]}}.

\bibitem{Babichev:2017guv}
E.~Babichev, C.~Charmousis, and A.~Lehébel, ``{Asymptotically flat black holes
  in Horndeski theory and beyond},''
  \href{http://dx.doi.org/10.1088/1475-7516/2017/04/027}{{\em JCAP} {\bf 04}
  (2017)  027}, \href{http://arxiv.org/abs/1702.01938}{{\tt arXiv:1702.01938
  [gr-qc]}}.

\bibitem{Lehebel:2017fag}
A.~Lehébel, E.~Babichev, and C.~Charmousis, ``{A no-hair theorem for stars in
  Horndeski theories},''
  \href{http://dx.doi.org/10.1088/1475-7516/2017/07/037}{{\em JCAP} {\bf 07}
  (2017)  037}, \href{http://arxiv.org/abs/1706.04989}{{\tt arXiv:1706.04989
  [gr-qc]}}.

\bibitem{Minamitsuji:2018vuw}
M.~Minamitsuji and H.~Motohashi, ``{Stealth Schwarzschild solution in shift
  symmetry breaking theories},''
  \href{http://dx.doi.org/10.1103/PhysRevD.98.084027}{{\em Phys. Rev. D} {\bf
  98} (2018) no.~8, 084027}, \href{http://arxiv.org/abs/1809.06611}{{\tt
  arXiv:1809.06611 [gr-qc]}}.

\bibitem{BenAchour:2019fdf}
J.~Ben~Achour, H.~Liu, and S.~Mukohyama, ``{Hairy black holes in DHOST
  theories: Exploring disformal transformation as a solution-generating
  method},'' \href{http://dx.doi.org/10.1088/1475-7516/2020/02/023}{{\em JCAP}
  {\bf 02} (2020)  023}, \href{http://arxiv.org/abs/1910.11017}{{\tt
  arXiv:1910.11017 [gr-qc]}}.

\bibitem{Minamitsuji:2019tet}
M.~Minamitsuji and J.~Edholm, ``{Black holes with a nonconstant kinetic term in
  degenerate higher-order scalar tensor theories},''
  \href{http://dx.doi.org/10.1103/PhysRevD.101.044034}{{\em Phys. Rev. D} {\bf
  101} (2020) no.~4, 044034}, \href{http://arxiv.org/abs/1912.01744}{{\tt
  arXiv:1912.01744 [gr-qc]}}.

\bibitem{Kobayashi:2012kh}
T.~Kobayashi, H.~Motohashi, and T.~Suyama, ``{Black hole perturbation in the
  most general scalar-tensor theory with second-order field equations I: the
  odd-parity sector},''
  \href{http://dx.doi.org/10.1103/PhysRevD.85.084025}{{\em Phys.\ Rev.\ D} {\bf
  85} (2012)  084025}, \href{http://arxiv.org/abs/1202.4893}{{\tt
  arXiv:1202.4893 [gr-qc]}}. [Erratum: Phys.Rev.D 96, 109903 (2017)].

\bibitem{Kobayashi:2014wsa}
T.~Kobayashi, H.~Motohashi, and T.~Suyama, ``{Black hole perturbation in the
  most general scalar-tensor theory with second-order field equations II: the
  even-parity sector},''
  \href{http://dx.doi.org/10.1103/PhysRevD.89.084042}{{\em Phys.\ Rev.\ D} {\bf
  89} (2014) no.~8, 084042}, \href{http://arxiv.org/abs/1402.6740}{{\tt
  arXiv:1402.6740 [gr-qc]}}.

\bibitem{Franciolini:2018uyq}
G.~Franciolini, L.~Hui, R.~Penco, L.~Santoni, and E.~Trincherini, ``{Effective
  Field Theory of Black Hole Quasinormal Modes in Scalar-Tensor Theories},''
  \href{http://dx.doi.org/10.1007/JHEP02(2019)127}{{\em JHEP} {\bf 02} (2019)
  127}, \href{http://arxiv.org/abs/1810.07706}{{\tt arXiv:1810.07706
  [hep-th]}}.

\bibitem{Babichev:2013cya}
E.~Babichev and C.~Charmousis, ``{Dressing a black hole with a time-dependent
  Galileon},'' \href{http://dx.doi.org/10.1007/JHEP08(2014)106}{{\em JHEP} {\bf
  08} (2014)  106}, \href{http://arxiv.org/abs/1312.3204}{{\tt arXiv:1312.3204
  [gr-qc]}}.

\bibitem{Kobayashi:2014eva}
T.~Kobayashi and N.~Tanahashi, ``{Exact black hole solutions in shift symmetric
  scalar--tensor theories},'' \href{http://dx.doi.org/10.1093/ptep/ptu096}{{\em
  PTEP} {\bf 2014} (2014)  073E02}, \href{http://arxiv.org/abs/1403.4364}{{\tt
  arXiv:1403.4364 [gr-qc]}}.

\bibitem{Babichev:2016kdt}
E.~Babichev and G.~Esposito-Farese, ``{Cosmological self-tuning and local
  solutions in generalized Horndeski theories},''
  \href{http://dx.doi.org/10.1103/PhysRevD.95.024020}{{\em Phys. Rev. D} {\bf
  95} (2017) no.~2, 024020}, \href{http://arxiv.org/abs/1609.09798}{{\tt
  arXiv:1609.09798 [gr-qc]}}.

\bibitem{Babichev:2017lmw}
E.~Babichev, C.~Charmousis, G.~Esposito-Farèse, and A.~Lehébel, ``{Stability
  of Black Holes and the Speed of Gravitational Waves within Self-Tuning
  Cosmological Models},''
  \href{http://dx.doi.org/10.1103/PhysRevLett.120.241101}{{\em Phys. Rev.
  Lett.} {\bf 120} (2018) no.~24, 241101},
  \href{http://arxiv.org/abs/1712.04398}{{\tt arXiv:1712.04398 [gr-qc]}}.

\bibitem{BenAchour:2018dap}
J.~Ben~Achour and H.~Liu, ``{Hairy Schwarzschild-(A)dS black hole solutions in
  degenerate higher order scalar-tensor theories beyond shift symmetry},''
  \href{http://dx.doi.org/10.1103/PhysRevD.99.064042}{{\em Phys. Rev. D} {\bf
  99} (2019) no.~6, 064042}, \href{http://arxiv.org/abs/1811.05369}{{\tt
  arXiv:1811.05369 [gr-qc]}}.

\bibitem{Motohashi:2019sen}
H.~Motohashi and M.~Minamitsuji, ``{Exact black hole solutions in
  shift-symmetric quadratic degenerate higher-order scalar-tensor theories},''
  \href{http://dx.doi.org/10.1103/PhysRevD.99.064040}{{\em Phys.\ Rev.\ D} {\bf
  99} (2019) no.~6, 064040}, \href{http://arxiv.org/abs/1901.04658}{{\tt
  arXiv:1901.04658 [gr-qc]}}.

\bibitem{Takahashi:2019oxz}
K.~Takahashi, H.~Motohashi, and M.~Minamitsuji, ``{Linear stability analysis of
  hairy black holes in quadratic degenerate higher-order scalar-tensor
  theories: Odd-parity perturbations},''
  \href{http://dx.doi.org/10.1103/PhysRevD.100.024041}{{\em Phys.\ Rev.\ D}
  {\bf 100} (2019) no.~2, 024041}, \href{http://arxiv.org/abs/1904.03554}{{\tt
  arXiv:1904.03554 [gr-qc]}}.

\bibitem{Minamitsuji:2019shy}
M.~Minamitsuji and J.~Edholm, ``{Black hole solutions in shift-symmetric
  degenerate higher-order scalar-tensor theories},''
  \href{http://dx.doi.org/10.1103/PhysRevD.100.044053}{{\em Phys. Rev. D} {\bf
  100} (2019) no.~4, 044053}, \href{http://arxiv.org/abs/1907.02072}{{\tt
  arXiv:1907.02072 [gr-qc]}}.

\bibitem{Khoury:2020aya}
J.~Khoury, M.~Trodden, and S.~S.~C. Wong, ``{Existence and instability of hairy
  black holes in shift-symmetric Horndeski theories},''
  \href{http://dx.doi.org/10.1088/1475-7516/2020/11/044}{{\em JCAP} {\bf 11}
  (2020)  044}, \href{http://arxiv.org/abs/2007.01320}{{\tt arXiv:2007.01320
  [astro-ph.CO]}}.

\bibitem{ArkaniHamed:2003uz}
N.~Arkani-Hamed, P.~Creminelli, S.~Mukohyama, and M.~Zaldarriaga, ``{Ghost
  inflation},'' \href{http://dx.doi.org/10.1088/1475-7516/2004/04/001}{{\em
  JCAP} {\bf 04} (2004)  001}, \href{http://arxiv.org/abs/hep-th/0312100}{{\tt
  arXiv:hep-th/0312100}}.

\bibitem{Mukohyama:2005rw}
S.~Mukohyama, ``{Black holes in the ghost condensate},''
  \href{http://dx.doi.org/10.1103/PhysRevD.71.104019}{{\em Phys.\ Rev.\ D} {\bf
  71} (2005)  104019}, \href{http://arxiv.org/abs/hep-th/0502189}{{\tt
  arXiv:hep-th/0502189}}.

\bibitem{Ogawa:2015pea}
H.~Ogawa, T.~Kobayashi, and T.~Suyama, ``{Instability of hairy black holes in
  shift-symmetric Horndeski theories},''
  \href{http://dx.doi.org/10.1103/PhysRevD.93.064078}{{\em Phys. Rev. D} {\bf
  93} (2016) no.~6, 064078}, \href{http://arxiv.org/abs/1510.07400}{{\tt
  arXiv:1510.07400 [gr-qc]}}.

\bibitem{Babichev:2018uiw}
E.~Babichev, C.~Charmousis, G.~Esposito-Farèse, and A.~Lehébel,
  ``{Hamiltonian unboundedness vs stability with an application to Horndeski
  theory},'' \href{http://dx.doi.org/10.1103/PhysRevD.98.104050}{{\em Phys.
  Rev. D} {\bf 98} (2018) no.~10, 104050},
  \href{http://arxiv.org/abs/1803.11444}{{\tt arXiv:1803.11444 [gr-qc]}}.

\bibitem{deRham:2019gha}
C.~de~Rham and J.~Zhang, ``{Perturbations of stealth black holes in degenerate
  higher-order scalar-tensor theories},''
  \href{http://dx.doi.org/10.1103/PhysRevD.100.124023}{{\em Phys.\ Rev.\ D}
  {\bf 100} (2019) no.~12, 124023}, \href{http://arxiv.org/abs/1907.00699}{{\tt
  arXiv:1907.00699 [hep-th]}}.

\bibitem{Takahashi:2021bml}
K.~Takahashi and H.~Motohashi, ``{Black hole perturbations in DHOST theories:
  master variables, gradient instability, and strong coupling},''
  \href{http://dx.doi.org/10.1088/1475-7516/2021/08/013}{{\em JCAP} {\bf 08}
  (2021)  013}, \href{http://arxiv.org/abs/2106.07128}{{\tt arXiv:2106.07128
  [gr-qc]}}.

\bibitem{Minamitsuji:2022mlv}
M.~Minamitsuji, K.~Takahashi, and S.~Tsujikawa, ``{Linear stability of black
  holes in shift-symmetric Horndeski theories with a time-independent scalar
  field},'' \href{http://dx.doi.org/10.1103/PhysRevD.105.104001}{{\em Phys.
  Rev. D} {\bf 105} (2022) no.~10, 104001},
  \href{http://arxiv.org/abs/2201.09687}{{\tt arXiv:2201.09687 [gr-qc]}}.

\bibitem{Minamitsuji:2022vbi}
M.~Minamitsuji, K.~Takahashi, and S.~Tsujikawa, ``{Linear stability of black
  holes with static scalar hair in full Horndeski theories: generic
  instabilities and surviving models},''
  \href{http://arxiv.org/abs/2204.13837}{{\tt arXiv:2204.13837 [gr-qc]}}.

\bibitem{Cheung:2007st}
C.~Cheung, P.~Creminelli, A.~L. Fitzpatrick, J.~Kaplan, and L.~Senatore, ``{The
  Effective Field Theory of Inflation},''
  \href{http://dx.doi.org/10.1088/1126-6708/2008/03/014}{{\em JHEP} {\bf 03}
  (2008)  014}, \href{http://arxiv.org/abs/0709.0293}{{\tt arXiv:0709.0293
  [hep-th]}}.

\bibitem{Finelli:2018upr}
B.~Finelli, G.~Goon, E.~Pajer, and L.~Santoni, ``{The Effective Theory of
  Shift-Symmetric Cosmologies},''
  \href{http://dx.doi.org/10.1088/1475-7516/2018/05/060}{{\em JCAP} {\bf 05}
  (2018)  060}, \href{http://arxiv.org/abs/1802.01580}{{\tt arXiv:1802.01580
  [hep-th]}}.

\bibitem{Mukohyama:2022enj}
S.~Mukohyama and V.~Yingcharoenrat, ``{Effective Field Theory of Black Hole
  Perturbations with Timelike Scalar Profile: Formulation},''
  \href{http://arxiv.org/abs/2204.00228}{{\tt arXiv:2204.00228 [hep-th]}}.

\bibitem{ArkaniHamed:2003uy}
N.~Arkani-Hamed, H.-C. Cheng, M.~A. Luty, and S.~Mukohyama, ``{Ghost
  condensation and a consistent infrared modification of gravity},''
  \href{http://dx.doi.org/10.1088/1126-6708/2004/05/074}{{\em JHEP} {\bf 05}
  (2004)  074}, \href{http://arxiv.org/abs/hep-th/0312099}{{\tt
  arXiv:hep-th/0312099}}.

\bibitem{Gleyzes:2013ooa}
J.~Gleyzes, D.~Langlois, F.~Piazza, and F.~Vernizzi, ``{Essential Building
  Blocks of Dark Energy},''
  \href{http://dx.doi.org/10.1088/1475-7516/2013/08/025}{{\em JCAP} {\bf 08}
  (2013)  025}, \href{http://arxiv.org/abs/1304.4840}{{\tt arXiv:1304.4840
  [hep-th]}}.

\bibitem{Akhoury:2008nn}
R.~Akhoury, C.~S. Gauthier, and A.~Vikman, ``{Stationary Configurations Imply
  Shift Symmetry: No Bondi Accretion for Quintessence / k-Essence},''
  \href{http://dx.doi.org/10.1088/1126-6708/2009/03/082}{{\em JHEP} {\bf 03}
  (2009)  082}, \href{http://arxiv.org/abs/0811.1620}{{\tt arXiv:0811.1620
  [astro-ph]}}.

\bibitem{TheLIGOScientific:2017qsa}
{\bf LIGO Scientific, Virgo} , B.~Abbott {\em et al.}, ``{GW170817: Observation
  of Gravitational Waves from a Binary Neutron Star Inspiral},''
  \href{http://dx.doi.org/10.1103/PhysRevLett.119.161101}{{\em Phys. Rev.
  Lett.} {\bf 119} (2017) no.~16, 161101},
  \href{http://arxiv.org/abs/1710.05832}{{\tt arXiv:1710.05832 [gr-qc]}}.

\bibitem{GBM:2017lvd}
{\bf LIGO Scientific, Virgo, Fermi GBM, INTEGRAL, IceCube, AstroSat Cadmium
  Zinc Telluride Imager Team, IPN, Insight-Hxmt, ANTARES, Swift, AGILE Team,
  1M2H Team, Dark Energy Camera GW-EM, DES, DLT40, GRAWITA, Fermi-LAT, ATCA,
  ASKAP, Las Cumbres Observatory Group, OzGrav, DWF (Deeper Wider Faster
  Program), AST3, CAASTRO, VINROUGE, MASTER, J-GEM, GROWTH, JAGWAR,
  CaltechNRAO, TTU-NRAO, NuSTAR, Pan-STARRS, MAXI Team, TZAC Consortium, KU,
  Nordic Optical Telescope, ePESSTO, GROND, Texas Tech University, SALT Group,
  TOROS, BOOTES, MWA, CALET, IKI-GW Follow-up, H.E.S.S., LOFAR, LWA, HAWC,
  Pierre Auger, ALMA, Euro VLBI Team, Pi of Sky, Chandra Team at McGill
  University, DFN, ATLAS Telescopes, High Time Resolution Universe Survey,
  RIMAS, RATIR, SKA South Africa/MeerKAT} , B.~Abbott {\em et al.},
  ``{Multi-messenger Observations of a Binary Neutron Star Merger},''
  \href{http://dx.doi.org/10.3847/2041-8213/aa91c9}{{\em Astrophys. J. Lett.}
  {\bf 848} (2017) no.~2, L12}, \href{http://arxiv.org/abs/1710.05833}{{\tt
  arXiv:1710.05833 [astro-ph.HE]}}.

\bibitem{Monitor:2017mdv}
{\bf LIGO Scientific, Virgo, Fermi-GBM, INTEGRAL} , B.~Abbott {\em et al.},
  ``{Gravitational Waves and Gamma-rays from a Binary Neutron Star Merger:
  GW170817 and GRB 170817A},''
  \href{http://dx.doi.org/10.3847/2041-8213/aa920c}{{\em Astrophys. J. Lett.}
  {\bf 848} (2017) no.~2, L13}, \href{http://arxiv.org/abs/1710.05834}{{\tt
  arXiv:1710.05834 [astro-ph.HE]}}.

\bibitem{Rubakov:2009np}
V.~A. Rubakov, ``{Harrison-Zeldovich spectrum from conformal invariance},''
  \href{http://dx.doi.org/10.1088/1475-7516/2009/09/030}{{\em JCAP} {\bf 09}
  (2009)  030}, \href{http://arxiv.org/abs/0906.3693}{{\tt arXiv:0906.3693
  [hep-th]}}.

\bibitem{Creminelli:2010ba}
P.~Creminelli, A.~Nicolis, and E.~Trincherini, ``{Galilean Genesis: An
  Alternative to inflation},''
  \href{http://dx.doi.org/10.1088/1475-7516/2010/11/021}{{\em JCAP} {\bf 11}
  (2010)  021}, \href{http://arxiv.org/abs/1007.0027}{{\tt arXiv:1007.0027
  [hep-th]}}.

\bibitem{Hinterbichler:2011qk}
K.~Hinterbichler and J.~Khoury, ``{The Pseudo-Conformal Universe: Scale
  Invariance from Spontaneous Breaking of Conformal Symmetry},''
  \href{http://dx.doi.org/10.1088/1475-7516/2012/04/023}{{\em JCAP} {\bf 04}
  (2012)  023}, \href{http://arxiv.org/abs/1106.1428}{{\tt arXiv:1106.1428
  [hep-th]}}.

\bibitem{Babichev:2018twg}
E.~Babichev, S.~Ramazanov, and A.~Vikman, ``{Recovering $P(X)$ from a canonical
  complex field},'' \href{http://dx.doi.org/10.1088/1475-7516/2018/11/023}{{\em
  JCAP} {\bf 11} (2018)  023}, \href{http://arxiv.org/abs/1807.10281}{{\tt
  arXiv:1807.10281 [gr-qc]}}.

\bibitem{Motohashi:2019ymr}
H.~Motohashi and S.~Mukohyama, ``{Weakly-coupled stealth solution in scordatura
  degenerate theory},''
  \href{http://dx.doi.org/10.1088/1475-7516/2020/01/030}{{\em JCAP} {\bf 01}
  (2020)  030}, \href{http://arxiv.org/abs/1912.00378}{{\tt arXiv:1912.00378
  [gr-qc]}}.

\bibitem{Gorji:2020bfl}
M.~A. Gorji, H.~Motohashi, and S.~Mukohyama, ``{Stealth dark energy in
  scordatura DHOST theory},''
  \href{http://dx.doi.org/10.1088/1475-7516/2021/03/081}{{\em JCAP} {\bf 03}
  (2021)  081}, \href{http://arxiv.org/abs/2009.11606}{{\tt arXiv:2009.11606
  [gr-qc]}}.

\bibitem{DeFelice:2022xvq}
A.~De~Felice, S.~Mukohyama, and K.~Takahashi, ``{Built-in scordatura in
  U-DHOST},'' \href{http://dx.doi.org/10.1103/PhysRevLett.129.031103}{{\em
  Phys. Rev. Lett.} {\bf 129} (2022)  031103},
  \href{http://arxiv.org/abs/2204.02032}{{\tt arXiv:2204.02032 [gr-qc]}}.

\bibitem{Hui:2021cpm}
L.~Hui, A.~Podo, L.~Santoni, and E.~Trincherini, ``{Effective Field Theory for
  the perturbations of a slowly rotating black hole},''
  \href{http://dx.doi.org/10.1007/JHEP12(2021)183}{{\em JHEP} {\bf 12} (2021)
  183}, \href{http://arxiv.org/abs/2111.02072}{{\tt arXiv:2111.02072
  [hep-th]}}.

\bibitem{Charmousis:2019vnf}
C.~Charmousis, M.~Crisostomi, R.~Gregory, and N.~Stergioulas, ``{Rotating Black
  Holes in Higher Order Gravity},''
  \href{http://dx.doi.org/10.1103/PhysRevD.100.084020}{{\em Phys. Rev. D} {\bf
  100} (2019) no.~8, 084020}, \href{http://arxiv.org/abs/1903.05519}{{\tt
  arXiv:1903.05519 [hep-th]}}.

\bibitem{BenAchour:2020fgy}
J.~Ben~Achour, H.~Liu, H.~Motohashi, S.~Mukohyama, and K.~Noui, ``{On rotating
  black holes in DHOST theories},''
  \href{http://dx.doi.org/10.1088/1475-7516/2020/11/001}{{\em JCAP} {\bf 11}
  (2020)  001}, \href{http://arxiv.org/abs/2006.07245}{{\tt arXiv:2006.07245
  [gr-qc]}}.

\bibitem{Mukohyama:2022skk}
S.~Mukohyama, K.~Takahashi, and V.~Yingcharoenrat, ``{Generalized Regge-Wheeler
  equation from Effective Field Theory of black hole perturbations with a
  timelike scalar profile},''
  \href{http://dx.doi.org/10.1088/1475-7516/2022/10/050}{{\em JCAP} {\bf 10}
  (2022)  050}, \href{http://arxiv.org/abs/2208.02943}{{\tt arXiv:2208.02943
  [gr-qc]}}.

\end{thebibliography}\endgroup

\end{document}